\documentclass[smallextended,natbib]{svjour3}

\usepackage{geometry}
\usepackage[colorlinks=true,linkcolor=blue,urlcolor=blue,citecolor=blue]{hyperref}
\usepackage[labelfont=bf,labelsep=space,justification=raggedright,singlelinecheck=false,font=small]{caption}
\usepackage{pgfplots,filecontents}
\usepackage{float}
\usepackage{siunitx}
\usepackage{booktabs,etoolbox}
\usepackage{multirow}
\usepackage{amssymb,amsmath}
\usepackage[ruled]{algorithm2e}
\usepackage{subcaption}

\usepgfplotslibrary{statistics}
\pgfplotsset{compat=1.5}

\smartqed

\begin{document}
	\title{Speeding up Memory-based Collaborative Filtering with Landmarks}

	\author{Gustavo R. Lima \and Carlos E. Mello \and Geraldo Zimbrao}

    \institute{Gustavo R. Lima \and Geraldo Zimbrao \at Federal University of Rio de Janeiro \\ \email{grlima@cos.ufrj.br; zimbrao@cos.ufrj.br}
            \and
            Carlos E. Mello \at Federal University of the State of Rio de Janeiro \\ \email{mello@uniriotec.br}
    }

	
	\maketitle
			
	\begin{abstract}
		Recommender systems play an important role in many scenarios where users are overwhelmed with too many choices to make. In this context, Collaborative Filtering (CF) arises by providing a simple and widely used approach for personalized recommendation. Memory-based CF algorithms mostly rely on similarities between pairs of users or items, which are posteriorly employed in classifiers like k-Nearest Neighbor (kNN) to generalize for unknown ratings. A major issue regarding this approach is to build the similarity matrix. Depending on the dimensionality of the rating matrix, the similarity computations may become computationally intractable. To overcome this issue, we propose to represent users by their distances to preselected users, namely landmarks. This procedure allows to drastically reduce the computational cost associated with the similarity matrix. We evaluated our proposal on two distinct distinguishing databases, and the results showed our method has consistently and considerably outperformed eight CF algorithms (including both memory-based and model-based) in terms of computational performance.
		
		\keywords{Recommender system \and Collaborative filtering \and Memory-based algorithms \and Landmarks \and Data reduction \and Dimensionality reduction \and Non-linear transformations}
	\end{abstract}

	\section{Introduction}

The continuously improving network technology and the exponential growth of social networks have been connecting the whole world, putting available a huge volume of content, media, goods, services, and many other different kinds of items on Internet \citep{pasinato2015active}. However, this phenomenon leads to the paradox of choice \citep{schwartz2004paradox}. It addresses the problem that people overwhelmed with too many choices tend to be more anxious, and eventually give up to proceed with the order.

To tackle this issue, a massive effort has been made towards the development of data mining methods for recommender systems \citep{ricci2011introduction}. This promising technology aims at helping users search and find items that are likely to be consumed, alleviating the burden of choice.

In this context, many recommender systems have been designed to provide users with suggested items in a personalized manner. A well-known and widely used approach for this kind of recommendation is Collaborative Filtering (CF) \citep{adomavicius2005toward}. It consists in considering the history of purchases and users' tastes to identify items that are likely to be acquired. In general, this data is represented by a rating matrix, where each row corresponds to a user, each column is assigned to an item, and each cell holds a rating given by the corresponding user and item. Thus, CF algorithms aim at predicting the missing ratings of the matrix, which are posteriorly used for personalized item recommendations.

CF algorithms may be divide into two main classes: \textit{memory-based} and \textit{model-based} algorithms. The former class uses k-Nearest Neighbors (kNN) methods for rating predictions, and therefore relies on computing similarities between pairs of users or items according to their ratings \citep{koren2011advances}. The latter class employs matrix factorization techniques so as to obtain an approximation of the rating matrix, in which the unknown cells are filled with rating predictions \citep{koren2009matrix}. Both memory-based and model-based algorithms provide advantages and disadvantages.

In this work, we are interested in memory-based algorithms. This class of CF algorithms remains widely used in many real systems due to its simplicity. It provides an elegant way for integrating information of users and items beyond the ratings for refining similarities \citep{shi2014collaborative}. In addition, memory-based CF algorithms allow \textit{online} recommendations, something required in many practical applications as data is arriving constantly, new users are signing up, and new products are being offered \citep{abernethy2007online}. So, incorporating such information in a \textit{online} fashion is very desired to make up-to-date predictions on the fly by avoiding to re-optimize from scratch with each new piece of data.

The major issue regarding to memory-based CF algorithms lies in its computational scalability associated with the growth of the rating matrix \citep{shi2014collaborative}. As users are often represented by vectors of items (\textit{i.e.} rows of the rating matrix), it turns out that the larger the number of items is, the higher the computational cost to compute similarities between users. Consequently, memory-based CF may become computationally intractable for a large number of users or items.

In this paper, we propose an alternative to improve the computational scalability of memory-based CF algorithms. Our proposal consists in representing users by their distances to preselected users, namely landmarks. Thus, instead of computing similarities between users represented by large vectors (often sparse) of ratings, our method calculates similarities through vectors of distances to fixed landmarks, obtaining an approximate similarity matrix for posterior rating predictions. As the number of landmarks required for a good approximation is mostly much smaller than the number of items, the proposed method drastically alleviates the cost associated with the similarity matrix computation.

The results show that our proposal consistently and considerably outperforms the evaluated CF algorithms (including both memory-based and model-based) in terms of computational performance. Interestingly, it achieves accuracy results better than the original memory-based CF algorithms with few landmarks. 

The main contributions of this work are the following:

\begin{itemize}
	\item A rating matrix reduction method to speed up memory-based CF algorithms.
	
	\item The proposal and investigation of 5 landmark selection strategies.
	
	\item An extensive comparison between our proposal and 8 CF algorithms, including both memory-based and model-based classes.
\end{itemize}

The work is organized in five sections, where this is the first one. Section 2 reviews the literature and presents the related work. Section 3 describes the recommendation problem definitions. It also introduces our proposal and presents the landmark selection strategies. Section 4 starts with the description of databases and metrics employed in experiments, follows by detailing the parameter tuning of the proposed method, and finishes by comparing our proposal against other CF algorithms. Finally, Section 5 points out conclusions and future work.

	\section{Related Work}

Collaborative Filtering (CF) approach consists in predicting whether a specific user would prefer an item rather than others based on ratings given by users \citep{adomavicius2005toward}. For this purpose, CF uses only a rating matrix $R$, where rows correspond to users, columns correspond to items, and each cell holds the rating value $r_{uv}$ given by user $u$ to item $v$.  Thus, the recommendation problem lies in predicting the missing ratings of $R$, which is often very sparse.

Interestingly, although there are many algorithms in Supervised Learning (SL) for data classification and regression, these are not properly suitable to CF, since ratings are not represented in a shared vector space $\mathbb{R}^{d}$. This happens because most users do not consume the same items by preventing their representation in the same vector space $\mathbb{R}^{d}$. Consequently, CF problem is slightly different from SL.

To overcome this issue, Braida et al. propose to build a vector space of latent factors to represent all item ratings given by users, and then apply SL techniques to predict unknown ratings. The authors use Singular Value Decomposition (SVD) to obtain user and item latent factors, and then build a vector space which contains all item ratings given by users. Their scheme consistently outperforms many state-of-the-art algorithms \citep{braida2015transforming}.

Sarwar et al. also apply SVD on the rating matrix to reduce its dimensionality and transform it in a new feature vector space. Thus, predictions are generated by operations between latent factor matrices of users and items \citep{sarwar2000application}.

Generally, dimensionality reduction techniques based on Matrix Factorization (MF) for CF are more efficient than other techniques, for instance Regularized SVD \citep{paterek2007improving}, Improved Regularized SVD \citep{paterek2007improving}, Probabilistic MF \citep{salakhutdinov2011probabilistic} and Bayesian Probabilistic MF \citep{salakhutdinov2008bayesian}. They have received great attention after Netflix Prize and are known as model-based CF algorithms \citep{breese1998empirical}.

In contrast, memory-based CF algorithms are an adapted k-Nearest Neighbors (kNN) method, in which similarity is computed considering only co-rated items between users, \textit{i.e.} the similarity between users are computed only for the vectors of co-rated items \citep{adomavicius2005toward}. Although model-based CF algorithms usually provide higher accuracy than the memory-based ones, the latter has been widely used \citep{beladev2015recommender,elbadrawy2015user,li2008ranking,pang2015cenknn,saleh2015promoting}. This is due to its simplicity in providing an elegant way for integrating information of users and items beyond the ratings for refining similarities \citep{shi2014collaborative}. Additionally, memory-based algorithms allow \textit{online} recommendations, making up-to-date predictions on the fly, which avoids to re-optimize from scratch with each new piece of data\citep{abernethy2007online}. For these reasons, many authors seek to improve memory-based CF accuracy and performance, for example in \citep{bobadilla2013similarity,gao2012novel,luo2013boosting}.

A well-known problem present in memory-based CF algorithms lies in applying distance functions to users for calculating their similarities, which are computationally expensive. Often, the algorithm runtime increases with the number of users/items, becoming prohibitive to apply it on very large databases. Furthermore, finding a sub-matrix of $R$ which contains all users and also is not empty might be impossible due to data sparsity, \textit{i.e.} it is difficult to find an item vector subspace in which all users are represented.

To tackle these issues, we propose a method to reduce the size rating matrix via landmarks. It consists in selecting $n$ users as landmarks, and then representing all users by their similarities to these landmarks. Thus, instead of representing users in item vector space, we propose to locate users in landmark vector space whose dimensionality is much smaller.

The landmark technique is useful to improve algorithm runtime and it was proposed by Silva and Tenenbaum in Multidimensional Scaling (MDS) context \citep{silva2002global}. In this case, the authors propose a Landmark MDS (LMDS) algorithm, which uses landmarks to reduce the computational costs of traditional MDS. LMDS builds a landmark set by selecting few observations from data -- the landmark set represents all observations. Then, it computes the similarity matrix for this set to obtain a suitable landmark representation in d-dimensional vector space. Finally, the other observations are mapped to this new space, considering their similarities to the landmarks \citep{de2004sparse}.

The main advantage of using LMDS instead of other techniques is to adjust accuracy and runtime. If one needs to decrease runtime, it is possible to sacrifice accuracy by reducing the size of the landmark set. Otherwise, if one needs to improve the algorithm's accuracy, it is also possible to increase the number of landmarks up to the database limit. Therefore, a good LMDS characteristic is to manage this trade off between runtime and accuracy \citep{platt2004fast}.

Lee and Choi \citep{lee2009landmark} argue that noise in database harms LMDS accuracy, and then propose an adaptation for this algorithm, namely Landmark MDS Ensemble (LMDS Ensemble). They propose applying LMDS to different data partitions, and then combine individual solutions in the same coordinate system. Their algorithm is less noise-sensitive but maintains computational performance of LMDS.

Another pitfall of landmark approach is to choose the most representative observation as landmarks, once the data representation depends on the similarity to these points. Several selection strategies are proposed in literature \citep{chen2006improved,chi2013selection,chi2014active,orsenigo2014improved,shi2016novel,shi2015landmark,silva2005selecting}, most of them related to select landmarks for Landmark Isomap, which is a nonlinear reduction method variation to improve scalability \citep{babaeian2015nonlinear,shang2011robust,silva2002global,sun2014ul}.

Finally, Hu et al. \citep{hu2009incremental} tackle the problem of applying Linear Discriminant Analysis (LDA) on databases where the number of samples is smaller than the data dimensionality. They propose joining MDS and LDA in an algorithm, named as Discriminant Multidimensional Mapping (DMM), and also employ landmarks in DMM (LDMM) to improve scalability and turn it feasible to very large databases.

	\section{Proposal}

We now present some basic definitions about memory-based Collaborative Filtering (CF) algorithms and discuss how it scales with the rating matrix size. Then, we follow by describing our proposal, which uses landmarks to improve the computational performance for computing the similarity matrix, and analyze its complexity. We also propose some selection strategies for the problem of choosing landmarks.

\subsection{Problem Definition}
\label{subsec:problem_definition}

For making predictions, memory-based CF algorithms consider similarities computed between pairs of users or pairs of items. Here, we assume that similarity is obtained between pair of users, namely used-based CF. The same can be done for pairs of items, namely item-based CF. 

User-based CF considers only the co-rated items to compute similarities between users, and predicts ratings for not yet rated items given a particular user \citep{adomavicius2005toward}. Thus, the items with highest predicted ratings are recommended.

In order to formally define the rating prediction problem, let $U$, $P$, $R$ be the set of users, the set of items and the rating matrix, respectively. Yet, let $V$ be the set of possible rating values in the recommender system. Thus, the rows of $R$ represent users and the columns represent items. If a user $u \in U$ rated an item $v \in P$ with the value $r_{uv} \in V$, then the cell at row $u$ and column $v$ of the matrix $R$ holds the value $r_{uv}$, otherwise it is empty. Consequently, the matrix $R$ dimension is $|U| \times |I|$ and, because most of the ratings are not provided, it is typically very sparse \citep{adomavicius2005toward}.

Let $P_{u}$ denote the item subset rated by a particular user $u$, and $P_{uu'} = P_{u} \cap P_{u'}$ the subset of items co-rated by users $u$ and $u'$. Note that, the recommender system aims at finding for a particular user $u$ the item $v \in P \setminus P_{u}$ to which the user $u$ is likely to be most interested. In other words, it estimates a function $f: U \times P \to V$ that predicts the rating $f(u,v)$ for a user $u$ and an item $v$. We denote the predicted rating by $\hat{r}_{uv}$ \citep{ricci2011introduction}.

To estimate this function, User-based CF employs a similarity measure $S: U \times U \to \mathbb{R}$ to determine the similarity $s_{uu'}$ between users $u$ and $u'$. Thus, the predicted rating $\hat{r}_{uv}$ is obtained with the k-Nearest Neighbors (kNN) rule given by \eqref{eq:predicted_rating}: 

\begin{equation}
\label{eq:predicted_rating}
\centering
\hat{r}_{uv} = \frac{\sum\limits_{u' \in U \setminus \{u\}}^{} s_{uu'}*(r_{u'v} - \bar{u'})}{\sum\limits_{u' \in U \setminus \{u\}}^{} s_{uu'}} + \bar{u},
\end{equation}
where $\bar{u}$ and $\bar{u'}$ represents the mean rating value of users $u$ and $u'$, respectively.

The most costly procedure in user-based CF is to compute the user-user similarity matrix. As the similarity measure must be applied to each pair of users in the system, which are represented by item ratings, a typical user-based CF must perform two nested loops, as one may see in algorithm \ref{alg:similarity}. These loops iterate over the user set and select a pair of users $u$ and $u'$ to compute their similarity. Thus, the algorithm complexity reaches $O(|U| \times |U| \times d)$, where $d$ indicates the similarity measure complexity, in case the number of items.

\begin{algorithm}
	\KwData{user set $U$, similarity measure $d$}
	\KwResult{user-user similarity matrix $S$}
	\For{$u \in U$}{
		\For{$u' \in U \setminus \{u\}$}{
			$S_{uu'} \gets d(u, u')$
		}
	}
	\caption{Algorithm to build user-user similarity matrix}
	\label{alg:similarity}
\end{algorithm}

Different similarity measures may be employed to build user-user similarity matrix and their complexity obviously depends on the number of operations performed.  To compute the similarity between users $u$ and $u'$, the measure must iterates over the item set. The algorithm \ref{alg:cosine_similarity} computes the Cosine similarity.

\begin{algorithm}
	\KwData{users $u$ and $u'$, item set $P$, rating matrix $R$}
	\KwResult{Cosine similarity $d_{uu'}$}
	
	$x, y, z \gets 0$
	
	\eIf{$|P_{uu'}| > 1$}{
		\For{$v \in P_{uu'}$}{
			$z \gets z  + r_{uv} * r_{u'v}$
			
			$x \gets x + r_{uv}^2$
			
			$y \gets y + r_{u'v}^2$
		}
		
		$d_{uu'} \gets z / (\sqrt{x} * \sqrt{y})$
	}
	{
		$d_{uu'} \gets -\infty$
	}
	\caption{The algorithm calculates the Cosine similarity between users $u$ and $u'$}
	\label{alg:cosine_similarity}
\end{algorithm}

Note that, algorithm \ref{alg:cosine_similarity} has a loop that iterates over co-rated items of users $u$ and $u'$. Therefore, user-based CF algorithm performs three nested loops and, consequently, their complexity is $O(|U| \times |U| \times |P|)$, which explains why its performance quickly decreases as the number of users increases.

\subsection{Building the New User Space}

In user-based CF, users are represented in item vector space, where components are the corresponding item ratings. Therefore, the user space dimensionality is $|P|$.

To improve user-based CF performance, we propose representing users in a space whose dimensionality is much smaller than the original one. The new vector space basis consists of preselected users from the rating matrix $R$, namely landmarks. The new user vector components are composed of the similarities to each corresponding landmark.

We select $n$ users from $R$, according to some criterion, like the number of item ratings. These $n$ users constitute the landmark set. Each landmark belongs to the original item vector space, with dimensionality $|P|$.

To build the new user space, one applies for each user $u \in U$ (including the landmarks) a non-linear transformation, which provides a smaller dimensional space. This transformation consists in computing the similarities between users and landmarks. These values forms the components of the new user vector representation. Therefore, to improve user-based CF performance, one must choose $n \ll |P|$. Additionally, the most representative landmarks should be preferred.

\begin{figure*}
	\includegraphics[width=\textwidth]{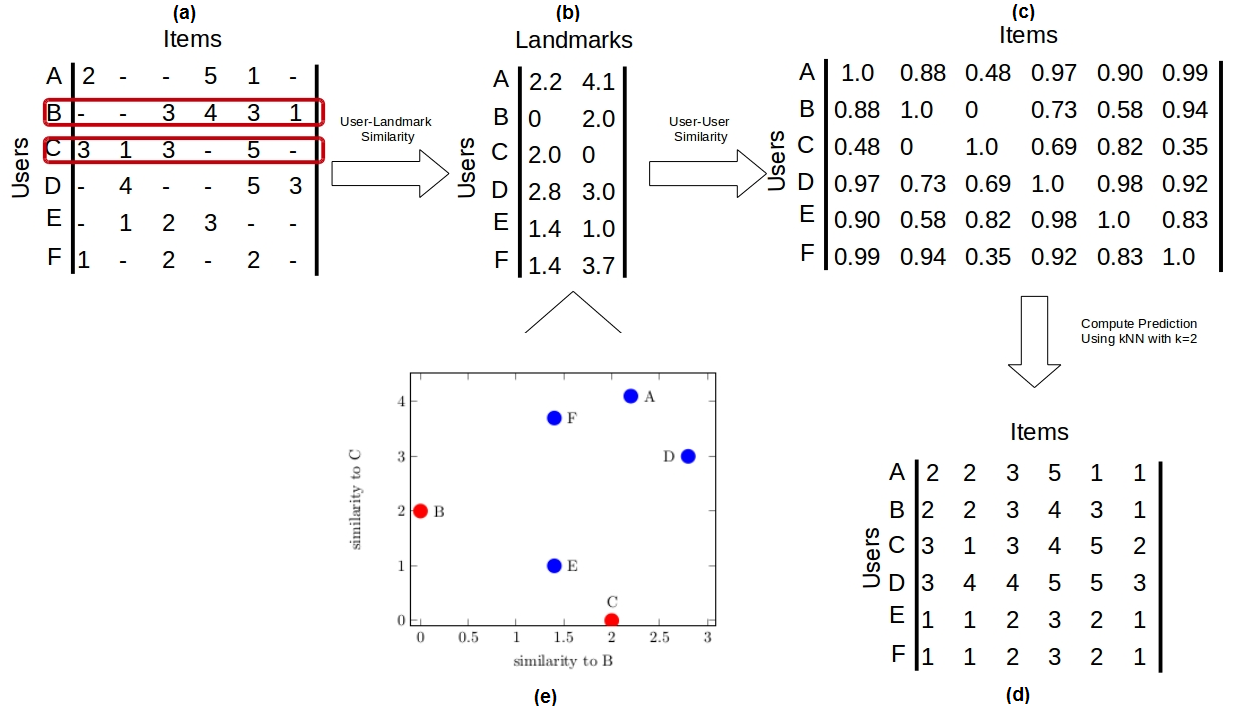}
	\caption{The figure shows the procedures to apply the proposed algorithm to a toy example. The algorithm input is the rating matrix illustrated in (a). The first step of algorithm is select landmarks, and then represent users in new vector space, as showed in (b) and (e). Then, similarities between users are computed in (c), and ratings predicted in (d).}
	\label{fig:toy_example}
\end{figure*}

\autoref{fig:toy_example} illustrates the proposed method for a toy example. In \hyperref[fig:toy_example]{\autoref{fig:toy_example}(a)}, the rating matrix $R$ contains the item ratings given by users $A$, $B$, $C$, $D$, $E$ and $F$. Missing ratings are indicated with `-'. Users $B$ and $C$ are selected as landmarks, considering the number of their given ratings (highlighted in red). 

In \hyperref[fig:toy_example]{\autoref{fig:toy_example}(b)}, the user-landmark similarity matrix is computed with Euclidean distance as a similarity measure. Note that, the rows of this matrix represent users, the two columns correspond to the landmarks $B$ and $C$, and its cells to corresponding similarities between the user and each landmark.

At this step, users are totally represented by their distance to landmarks. In other words, the landmarks constitute the basis of the new vector space and each user is positioned on according to the landmarks. 

An advantage of the new representation may be seen in the toy example: users $A$ and $D$ are co-rated at only one item, and therefore computing their similarity would produce a low accurate value. However, the new space locates users $A$ and $D$ according to their distance to landmarks $B$ and $C$, which is here computed with more than one co-rated item. Therefore, users $A$ and $B$ are positioned near each other, as may be seen on \hyperref[fig:toy_example]{\autoref{fig:toy_example}(e)}.

Once computed the new reduced user space, the next step is to compute the user-user similarity matrix based on this new representation. In \hyperref[fig:toy_example]{\autoref{fig:toy_example}(c)}, one applied the Cosine similarity measure. Finally, it predicts the missing ratings with kNN approach, using k=2, as shown in \hyperref[fig:toy_example]{\autoref{fig:toy_example}(d)}.

The computation of the user-user similarity matrix via landmarks is presented in algorithm \ref{alg:landmarks}. The first part of this algorithm consists in selecting $n$ landmarks based on some criterion defined by the function $selectLandmarks$. After choosing the landmarks, users are represented in the landmark vector space by calculating their similarities to landmarks -- the similarity measure $d_{1}$. Finally, the last step builds the user-user similarity matrix for the new landmark representation -- the similarity measure $d_{2}$.

\begin{algorithm}
	\KwData{user set $U$, number of landmarks $n$, user-landmark similarity measure $d_{1}$, user-user similarity measure $d_{2}$}
	\KwResult{user-user similarity matrix $S$}
	
	$L \gets selectLandmarks(n)$
	
	$H \gets zeroMatrix(|U|,n)$
	
	\For{$u \in U$}{
		\For{$l \in L$}{
			$H_{ul} \gets d_{1}(u,l)$
		}
	}
	
	\For{$u \in U$}{
		\For{$u' \in U \setminus \{u\}$}{
			$S_{uu'} \gets d_{2}(u,u')$
		}
	}
	\caption{Algorithm to build user-user similarity matrix using landmarks.}
	\label{alg:landmarks}
\end{algorithm}

The difference between similarity measures $d_{1}$ and $d_{2}$ is that the former considers user ratings to compute similarities, while the latter considers user-landmark similarities to achieve its purpose. Algorithms \ref{alg:cosine_similarity} and \ref{alg:cosine_landmarks_similarity} illustrates both cases, respectively.

\begin{algorithm}
	\KwData{users $u$ and $u'$, landmark set $L$, user-landmark similarity matrix $H$}
	\KwData{Cosine similarity $S_{uu'}$}
	$x, y, z \gets 0$
	
	\For{$l \in L$}{
		\If{$H_{ul}\ !=\ -\infty$ and $H_{u'l}\ !=\ -\infty$}{
			$z \gets z  + H_{ul} * H_{u'l}$
			
			$x \gets x + H_{ul}^2$
			
			$y \gets y + H_{u'l}^2$
		}
	}
	
	$S_{uu'} = z / (\sqrt{x} * \sqrt{y})$
	\caption{Algorithm to compute Cosine similarity between user $u$ and $u'$ which are represented in landmark vector space}
	\label{alg:cosine_landmarks_similarity}
\end{algorithm}

\subsection{Choosing Landmarks}

The choice of landmarks is a critical task, once it directly affects the resulting vector space, and, consequently, influences the accuracy of user-based CF. To investigate this effect we propose five landmark selection strategies, described as follows:
\begin{itemize}
	\item \textit{Random} - $n$ users are randomly chosen as landmarks.
	
	\item \textit{Dist. of Ratings} - randomly chooses $n$ users as landmarks by considering the distribution of ratings, \textit{i.e.} users with more ratings are more likely to be selected.
	
	\item \textit{Coresets} - chooses at random $n$ landmark candidates taking into account the rating distribution. Then, it computes the user similarity to these candidates and removes half of the most similar users. From the remaining users, $n$ new landmark candidates are again randomly chosen and the half most similar users to the candidates are removed. The algorithm proceeds until no remaining user exists in database. This strategy is based on \textit{coresets}\citep{feldman2011scalable}.
	
	\item \textit{Coresets Random} - does similar to \textit{Coresets}, but without considering the rating distribution, instead it just samples users uniformly at random.

	\item \textit{Popularity} - ranks users/items in descend order by their number of ratings and selects the first $n$ users as landmarks. 
\end{itemize}

The proposed landmark selection strategies have underlying different criterion. Three of them -- \textit{Dist. of Ratings}, \textit{Coresets} and \textit{Popularity} -- consider the number of ratings as criterion to select landmarks. Thus, we expect to select more representative landmarks compared with the other two strategies -- \textit{Random} and \textit{Coresets Random}.

Additionally, \textit{Random} and \textit{Dist. of Ratings} are both the simplest, and consequently the fastest strategies, since landmarks are selected with no data preprocessing. \textit{Popularity} is an intermediate case, as one needs to sort users/items by their number of ratings, which requires more computations than random sampling.

Finally, \textit{Coresets} and \textit{Coresets Random} are the most complex strategies. One should compute user similarities to $n$ landmark candidates and remove the half most similar users from the entire set. This process proceeds until no user remains. Thus, both strategies require more computations than \textit{Popularity}.

\subsection{Complexity Analysis}

In our proposal, users are represented in the landmark vector space. Once landmarks are chosen, the proposed algorithm performs three nested loops to compute the user-landmark similarity matrix. The first loop iterates over all users and the second one over $n$ landmarks. Then, for each user $u$ and landmark $l$, the loop inside the similarity measure $d_{1}$ iterates over the co-rated items of $u$ and $l$ and computes the corresponding similarity value. Therefore, to build the user-landmark similarity matrix, one requires $O(|U| \times n \times |P|)$ steps.

To compute user-user similarity matrix, the algorithm performs three nested loops. The first two iterate over all users and the one inside the similarity measure $d_{2}$ iterates over $n$ landmarks. Thus, it results in $O(|U| \times |U| \times n)$ steps.

Accordingly, the proposed algorithm complexity is $O(|U| \times n \times |P| + |U| \times |U| \times n) = O(|U| \times n \times (|P| + |U|))$.

Note that, this complexity becomes smaller as the number $n$ of landmarks decreases. By comparing this result with the original complexity of user-based CF, $O(|U| \times |U| \times |P|)$, it turns out that $O(|U| \times n \times (|P| + |U|)) \leq O(|U| \times |U| \times |P|)$ if $n \leq \frac{|U| \times |P|}{|U| + |P|}$, what is a very realistic assumption.
	\section{Experiments}

\subsection{Methodology}

In order to analyze the proposed method performance, we conduct experiments on two well-known databases: MovieLens \citep{harper2016movielens,miller2003movielens} and Netflix \citep{bennett2007netflix}. Our objective was twofold: (1) parameter investigation and settings -- to investigate the functioning of the proposed method with regards to its parameter settings such as the number of landmarks, the similarity measure to build the landmark-user matrix, and the landmark selection strategy; and, (2) comparative analysis -- to compare the proposal with different state-of-the-art algorithms.

In (1), the parameter investigation and settings, we start by varying the number of landmarks from 10 up to 100 for each selection strategy. The idea is to evaluate the algorithm prediction accuracy with different parameter settings and how these may be affected. Still, we also evaluate different similarity measures either to build the landmark-user/item matrix or the similarity matrix for rating predictions. The similarity measures were Euclidean, Cosine, and Pearson. 

In (2), the comparative analysis, there were compared several algorithms from both memory-based and model-based algorithms. These are listed as follows:

\begin{enumerate}
\item Memory-based algorithms:
    \begin{enumerate}
	\item k-Nearest Neighbor (kNN) with Euclidean \citep{adomavicius2005toward};
	\item kNN with Cosine \citep{adomavicius2005toward};
	\item kNN with Pearson \citep{adomavicius2005toward};
    \end{enumerate}
\item Model-based algorithms:
	\begin{enumerate}
	\item Regularized Singular Value Decomposition (SVD) \citep{paterek2007improving};
	\item Improved Regularized Singular Value Decomposition \citep{paterek2007improving};
	\item Probability Matrix Factorization (MF) \citep{salakhutdinov2011probabilistic};
	\item Bayesian Probability Matrix Factorization \citep{salakhutdinov2008bayesian}; and
	\item SVD++ \citep{koren2008factorization,koren2009matrix}.
    \end{enumerate}
\end{enumerate}

All experiments were carried out with 10-fold cross validation. There were considered the mean absolute error (MAE) to measure accuracy of rating predictions and the runtime in seconds to assess computational performance \citep{herlocker2004evaluating}.

\subsection{Data sets}
\label{subsec:databases}

In order to pursue a fair evaluation, we consider two different data set sizes from MoviesLens and Netflix databases. In the former database, there were already two available data sets: MovieLens100k and MovieLens1M; each one with one hundred thousand (100k) and one million (1M) ratings, respectively\citep{herlocker1999algorithmic}. In the latter database, Netflix, it was necessary to cut out the original database by generating data sets with amounts of ratings equal to the two MovieLens data sets. Thus, there were extracted 100k and 1M ratings in chronological recording order from the original database, obtaining two data sets: Netflix100k and Netflix1M, respectively. The idea of these cuts was to preserve temporal characteristics. \autoref{tab:datasets-characteristics} shows the number of users and items, and the sparsity of the rating matrix in the four data sets.

\begin{table}[]
    \fontsize{8pt}{10pt}
    \selectfont
    \caption{Data set characteristics.}
    \centering
    \label{my-label}
    \begin{tabular}{l l l l l}
	\hline\noalign{\smallskip}
                  & \multicolumn{1}{l}{\#ratings} & \multicolumn{1}{l}{\#users} & \multicolumn{1}{l}{\#items} & \multicolumn{1}{l}{sparsity(\%)} \\
    \noalign{\smallskip}\hline\noalign{\smallskip}
    MovieLens100k & 100k                           & 943                          & 1,682                        & 6.3                               \\
    Netflix100k   & 100k                           & 1,490                        & 2,380                        & 2.82                              \\
    MovieLens1M   & 1M                             & 6,040                        & 3,952                        & 4.19                              \\
    Netflix1M   & 1M                             & 8,782                        & 4,577                        & 2.48                              \\
    \noalign{\smallskip}\hline
    \end{tabular}
    \label{tab:datasets-characteristics}
\end{table}

\subsection{Parameter investigation and settings}
We here aim to analyze and discuss how the proposed algorithm works with different parameter values.

\subsubsection{Accuracy Analysis}

The graphs in \autoref{fig:accuracy_per_landmarks_100k} and \ref{fig:accuracy_per_landmarks_1M} present MAE per number of landmarks on the data sets of 100k and 1M ratings, respectively. The set of landmarks was varied from 10 up to 100, and, at each 10 landmarks, we compute MAE. This procedure was conducted for five landmark selection strategies: \text{Random}, \text{Dist. of Ratings}, \text{Coresets}, \text{Coresets Random}, and \text{Popularity}. In addition, these results were compared with the original version of the memory-based CF algorithm, \textit{i.e.} user/item-based Collaborative Filtering (CF) with Cosine similarity, namely \textit{baseline} algorithm.

In this experiment, the proposed method employed Euclidean distance to build the user-landmark matrix, and then obtained the user-user similarity matrix by Cosine distance, for reducing the dimensionality of user-based CF. Analogously, the same settings were adopted for item-based CF.

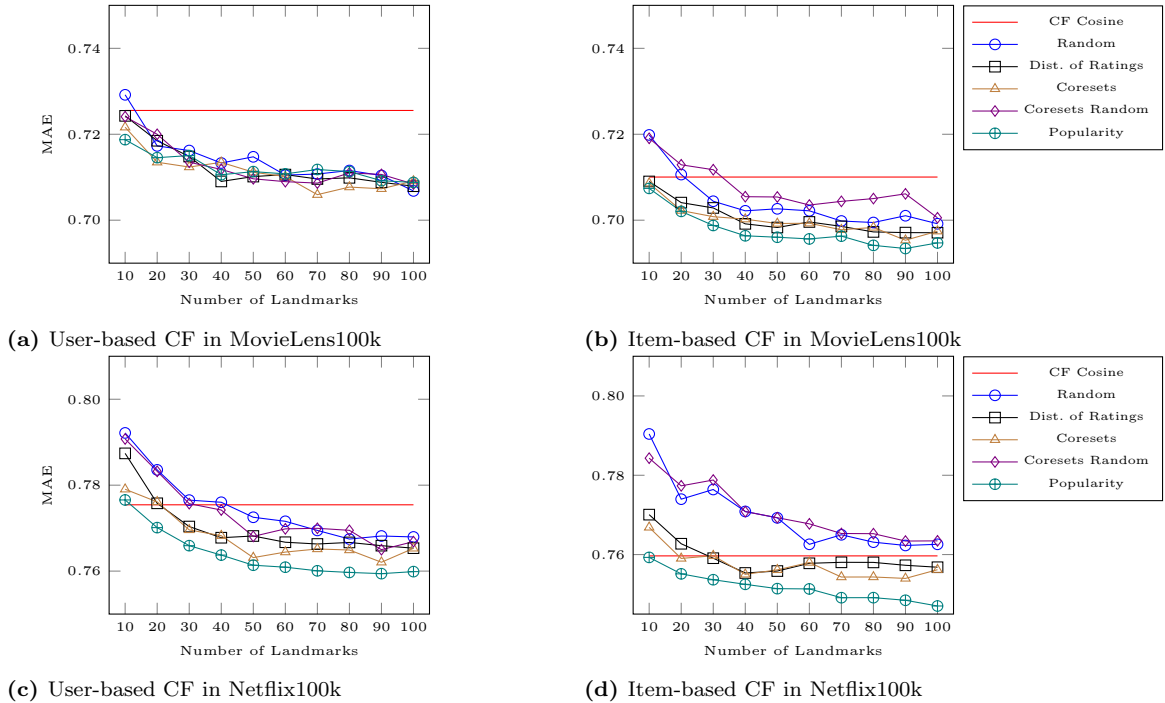
\begin{figure*}
	\begin{subfigure}{0.4\textwidth}
	    \centering
	    
	    \begin{tikzpicture}
	    \begin{axis} [xtick={10,20,30,40,50,60,70,80,90,100},
	        ymin=0.69, ymax=0.750,
	        xmin=5, xmax=105,
	        x tick label style={font=\tiny},
	        y tick label style={/pgf/number format/.cd,
	            fixed,
	            fixed zerofill,
	            precision=2,
	            /tikz/.cd,
	            font=\tiny},
	        xlabel=Number of Landmarks,
	        xlabel style={font=\tiny},
	        ylabel=MAE,
	        ylabel style={font=\tiny},
	        legend style={font=\fontsize{4.5}{4.5}\selectfont},
	        height=5cm]
	        
	        \addplot[red, mark=\empty] table [x index=0, y index=7] {ML_100k_user-based_mae.dat};
	        
	        \addplot[blue, mark=o, solid] table [x index=0, y index=1] {ML_100k_user-based_mae.dat};
	        
	        \addplot[black, mark=square, solid] table [x index=0, y index=2] {ML_100k_user-based_mae.dat};
	        
	        \addplot[brown, mark=triangle, solid] table [x index=0, y index=3] {ML_100k_user-based_mae.dat};
	        
	        \addplot[violet, mark=diamond, solid] table [x index=0, y index=4] {ML_100k_user-based_mae.dat};
	        
	        \addplot[teal, mark=oplus, solid] table [x index=0, y index=5] {ML_100k_user-based_mae.dat};
	    
	        \end{axis}
	    \end{tikzpicture}
	    
	    \caption{User-based CF in MovieLens100k}
	    \label{fig:landmarks_ml_100k_user-based_mae}
	\end{subfigure}
	~~~~~~~~~~~~~~
	\begin{subfigure}{0.4\textwidth}
	    \centering
	    
	    \begin{tikzpicture}
	    \begin{axis} [xtick={10,20,30,40,50,60,70,80,90,100},
	        ymin=0.690, ymax=0.750,
	        xmin=5, xmax=105,
	        x tick label style={font=\tiny},
	        y tick label style={/pgf/number format/.cd,
	            fixed,
	            fixed zerofill,
	            precision=2,
	            /tikz/.cd,
	            font=\tiny},
	        xlabel=Number of Landmarks,
	        xlabel style={font=\tiny},
	        ylabel style={font=\tiny},
	        legend style={font=\fontsize{4.5}{4.5}\selectfont},
	        legend pos=outer north east,
	        height=5cm]
	        
	        \addplot[red, mark=\empty] table [x index=0, y index=7] {ML_100k_item-based_mae.dat};
	        \addlegendentry{CF Cosine}
	        
	        \addplot[blue, mark=o, solid] table [x index=0, y index=1] {ML_100k_item-based_mae.dat};
	        \addlegendentry{Random}
	        
	        \addplot[black, mark=square, solid] table [x index=0, y index=2] {ML_100k_item-based_mae.dat};
	        \addlegendentry{Dist. of Ratings}
	        
	        \addplot[brown, mark=triangle, solid] table [x index=0, y index=3] {ML_100k_item-based_mae.dat};    \addlegendentry{Coresets}
	        
	        \addplot[violet, mark=diamond, solid] table [x index=0, y index=4] {ML_100k_item-based_mae.dat};
	        \addlegendentry{Coresets Random}
	        
	        \addplot[teal, mark=oplus, solid] table [x index=0, y index=5] {ML_100k_item-based_mae.dat};
	        \addlegendentry{Popularity}
	    
	        \end{axis}
	    \end{tikzpicture}
	    
	    \caption{Item-based CF in MovieLens100k}
	    \label{fig:landmarks_ml_100k_item-based_mae}
	\end{subfigure}
	~
	
	\begin{subfigure}{0.4\textwidth}
	    \centering
	    
	    \begin{tikzpicture}
	    \begin{axis} [xtick={10,20,30,40,50,60,70,80,90,100},
	        ymin=0.750, ymax=0.810,
	        xmin=5, xmax=105,
	        x tick label style={font=\tiny},
	        y tick label style={/pgf/number format/.cd,
	            fixed,
	            fixed zerofill,
	            precision=2,
	            /tikz/.cd,
	            font=\tiny},
	        xlabel=Number of Landmarks,
	        xlabel style={font=\tiny},
	        ylabel=MAE,
	        ylabel style={font=\tiny},
	        legend style={font=\fontsize{4.5}{4.5}\selectfont},
	        height=5cm]
	    
    	    \addplot[red, mark=\empty] table [x index=0, y index=7] {Netflix_100k_user-based_mae.dat};
    	
    	    \addplot[blue, mark=o, solid] table [x index=0, y index=1] {Netflix_100k_user-based_mae.dat};
    	    
    	    \addplot[black, mark=square, solid] table [x index=0, y index=2] {Netflix_100k_user-based_mae.dat};
    	    
    	    \addplot[brown, mark=triangle, solid] table [x index=0, y index=3] {Netflix_100k_user-based_mae.dat};
    	    
    	    \addplot[violet, mark=diamond, solid] table [x index=0, y index=4] {Netflix_100k_user-based_mae.dat};
    	    
    	    \addplot[teal, mark=oplus, solid] table [x index=0, y index=5] {Netflix_100k_user-based_mae.dat};
    	
    	    \end{axis}
    	\end{tikzpicture}
	    
	    \caption{User-based CF in Netflix100k}
	    \label{fig:landmarks_netflix_100k_user-based_mae}
	\end{subfigure}
	~~~~~~~~~~~~~~
	\begin{subfigure}{0.4\textwidth}
	    \centering
	    
	    \begin{tikzpicture}
	    \begin{axis} [xtick={10,20,30,40,50,60,70,80,90,100},
	        ymin=0.745, ymax=0.810,
	        xmin=5, xmax=105,
	        x tick label style={font=\tiny},
	        y tick label style={/pgf/number format/.cd,
	            fixed,
	            fixed zerofill,
	            precision=2,
	            /tikz/.cd,
	            font=\tiny},
	        xlabel=Number of Landmarks,
	        xlabel style={font=\tiny},
	        ylabel style={font=\tiny},
	        legend style={font=\fontsize{4.5}{4.5}\selectfont},
	        legend pos=outer north east,
	        height=5cm]
	    
    	    \addplot[red, mark=\empty] table [x index=0, y index=7] {Netflix_100k_item-based_mae.dat};
    	    \addlegendentry{CF Cosine}
    	
    	    \addplot[blue, mark=o, solid] table [x index=0, y index=1] {Netflix_100k_item-based_mae.dat};
    	    \addlegendentry{Random}
    	    
    	    \addplot[black, mark=square, solid] table [x index=0, y index=2] {Netflix_100k_item-based_mae.dat};
    	    \addlegendentry{Dist. of Ratings}
    	    
    	    \addplot[brown, mark=triangle, solid] table [x index=0, y index=3] {Netflix_100k_item-based_mae.dat};    \addlegendentry{Coresets}
    	    
    	    \addplot[violet, mark=diamond, solid] table [x index=0, y index=4] {Netflix_100k_item-based_mae.dat};
    	    \addlegendentry{Coresets Random}
    	    
    	    \addplot[teal, mark=oplus, solid] table [x index=0, y index=5] {Netflix_100k_item-based_mae.dat};
    	    \addlegendentry{Popularity}
    	
    	    \end{axis}
    	\end{tikzpicture}
	    
	    \caption{Item-based CF in Netflix100k}
	    \label{fig:landmarks_netflix_100k_item-based_mae}
	\end{subfigure}
	
	\caption{The figure shows how MAE behaves as landmark increases for CF in 100k rating data sets.}
	\label{fig:accuracy_per_landmarks_100k}
\end{figure*}
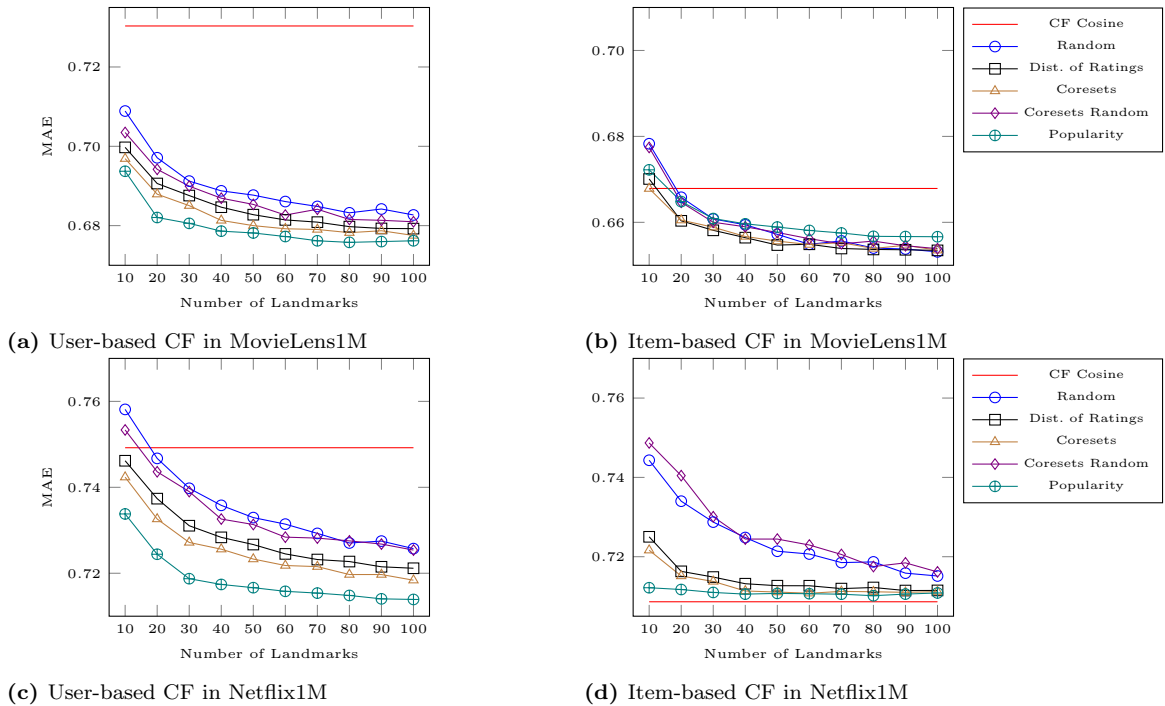
\begin{figure*}
	\begin{subfigure}{0.4\textwidth}
	    \centering
	    
	    \begin{tikzpicture}
	    \begin{axis} [xtick={10,20,30,40,50,60,70,80,90,100},
	        ymin=0.670, ymax=0.735,
	        xmin=5, xmax=105,
	        x tick label style={font=\tiny},
	        y tick label style={/pgf/number format/.cd,
	            fixed,
	            fixed zerofill,
	            precision=2,
	            /tikz/.cd,
	            font=\tiny},
	        xlabel=Number of Landmarks,
	        xlabel style={font=\tiny},
	        ylabel=MAE,
	        ylabel style={font=\tiny},
	        legend style={font=\fontsize{4.5}{4.5}\selectfont},
	        height=5cm]
	        
    	        \addplot[red, mark=\empty] table [x index=0, y index=7] {ML_1M_user-based_mae.dat};
    	        
    	        \addplot[blue, mark=o, solid] table [x index=0, y index=1] {ML_1M_user-based_mae.dat};
    	        
    	        \addplot[black, mark=square, solid] table [x index=0, y index=2] {ML_1M_user-based_mae.dat};
    	        
    	        \addplot[brown, mark=triangle, solid] table [x index=0, y index=3] {ML_1M_user-based_mae.dat};
    	        
    	        \addplot[violet, mark=diamond, solid] table [x index=0, y index=4] {ML_1M_user-based_mae.dat};
    	        
    	        \addplot[teal, mark=oplus, solid] table [x index=0, y index=5] {ML_1M_user-based_mae.dat};
    	    
    	        \end{axis}
    	    \end{tikzpicture}
	    
	    \caption{User-based CF in MovieLens1M}
	    \label{fig:landmarks_ml_1M_user-based_mae}
	\end{subfigure}
	~~~~~~~~~~~~~~
	\begin{subfigure}{0.4\textwidth}
	    \centering
	    
	    \begin{tikzpicture}
	    \begin{axis} [xtick={10,20,30,40,50,60,70,80,90,100},
	        ymin=0.650, ymax=0.71,
	        xmin=5, xmax=105,
	        x tick label style={font=\tiny},
	        y tick label style={/pgf/number format/.cd,
	            fixed,
	            fixed zerofill,
	            precision=2,
	            /tikz/.cd,
	            font=\tiny},
	        xlabel=Number of Landmarks,
	        xlabel style={font=\tiny},
	        ylabel style={font=\tiny},
	        legend style={font=\fontsize{4.5}{4.5}\selectfont},
	        legend pos=outer north east,
	        height=5cm]
	        
	        \addplot[red, mark=\empty] table [x index=0, y index=7] {ML_1M_item-based_mae.dat};
	        \addlegendentry{CF Cosine}
	        
	        \addplot[blue, mark=o, solid] table [x index=0, y index=1] {ML_1M_item-based_mae.dat};
	        \addlegendentry{Random}
	        
	        \addplot[black, mark=square, solid] table [x index=0, y index=2] {ML_1M_item-based_mae.dat};
	        \addlegendentry{Dist. of Ratings}
	        
	        \addplot[brown, mark=triangle, solid] table [x index=0, y index=3] {ML_1M_item-based_mae.dat};
	        \addlegendentry{Coresets}
	        
	        \addplot[violet, mark=diamond, solid] table [x index=0, y index=4] {ML_1M_item-based_mae.dat};
	        \addlegendentry{Coresets Random}
	        
	        \addplot[teal, mark=oplus, solid] table [x index=0, y index=5] {ML_1M_item-based_mae.dat};
	        \addlegendentry{Popularity}
	    
	        \end{axis}
	    \end{tikzpicture}
	    
	    \caption{Item-based CF in MovieLens1M}
	    \label{fig:landmarks_ml_1M_item-based_mae}
	\end{subfigure}
	~
	
	\begin{subfigure}{0.4\textwidth}
	    \centering
	    
	    \begin{tikzpicture}
	    \begin{axis} [xtick={10,20,30,40,50,60,70,80,90,100},
	        ymin=0.710, ymax=0.770,
	        xmin=5, xmax=105,
	        x tick label style={font=\tiny},
	        y tick label style={/pgf/number format/.cd,
	            fixed,
	            fixed zerofill,
	            precision=2,
	            /tikz/.cd,
	            font=\tiny},
	        xlabel=Number of Landmarks,
	        xlabel style={font=\tiny},
	        ylabel=MAE,
	        ylabel style={font=\tiny},
	        legend style={font=\fontsize{4.5}{4.5}\selectfont},
	        height=5cm]
		    
		    \addplot[red, mark=\empty] table [x index=0, y index=7] {Netflix_1M_user-based_mae.dat};
		
		    \addplot[blue, mark=o, solid] table [x index=0, y index=1] {Netflix_1M_user-based_mae.dat};
		    
		    \addplot[black, mark=square, solid] table [x index=0, y index=2] {Netflix_1M_user-based_mae.dat};
		    
		    \addplot[brown, mark=triangle, solid] table [x index=0, y index=3] {Netflix_1M_user-based_mae.dat};
		    
		    \addplot[violet, mark=diamond, solid] table [x index=0, y index=4] {Netflix_1M_user-based_mae.dat};
		    
		    \addplot[teal, mark=oplus, solid] table [x index=0, y index=5] {Netflix_1M_user-based_mae.dat};
		
		    \end{axis}
		\end{tikzpicture}
	    
	    \caption{User-based CF in Netflix1M}
	    \label{fig:landmarks_netflix_1M_user-based_mae}
	\end{subfigure}
	~~~~~~~~~~~~~~
	\begin{subfigure}{0.4\textwidth}
	    \centering
	    
	    \begin{tikzpicture}
	    \begin{axis} [xtick={10,20,30,40,50,60,70,80,90,100},
	        ymin=0.705, ymax=0.770,
	        xmin=5, xmax=105,
	        x tick label style={font=\tiny},
	        y tick label style={/pgf/number format/.cd,
	            fixed,
	            fixed zerofill,
	            precision=2,
	            /tikz/.cd,
	            font=\tiny},
	        xlabel=Number of Landmarks,
	        xlabel style={font=\tiny},
	        ylabel style={font=\tiny},
	        legend style={font=\fontsize{4.5}{4.5}\selectfont},
	        legend pos=outer north east,
	        height=5cm]
	    
    	    \addplot[red, mark=\empty] table [x index=0, y index=7] {Netflix_1M_item-based_mae.dat};
    	    \addlegendentry{CF Cosine}
    	
    	    \addplot[blue, mark=o, solid] table [x index=0, y index=1] {Netflix_1M_item-based_mae.dat};
    	    \addlegendentry{Random}
    	    
    	    \addplot[black, mark=square, solid] table [x index=0, y index=2] {Netflix_1M_item-based_mae.dat};
    	    \addlegendentry{Dist. of Ratings}
    	    
    	    \addplot[brown, mark=triangle, solid] table [x index=0, y index=3] {Netflix_1M_item-based_mae.dat};
    	    \addlegendentry{Coresets}
    	    
    	    \addplot[violet, mark=diamond, solid] table [x index=0, y index=4] {Netflix_1M_item-based_mae.dat};
    	    \addlegendentry{Coresets Random}
    	    
    	    \addplot[teal, mark=oplus, solid] table [x index=0, y index=5] {Netflix_1M_item-based_mae.dat};
    	    \addlegendentry{Popularity}
    	
    	    \end{axis}
    	\end{tikzpicture}
	    
	    \caption{Item-based CF in Netflix1M}
	    \label{fig:landmarks_netflix_1M_item-based_mae}
	\end{subfigure}
	
	\caption{The figure shows how MAE behaves as landmark increases for CF in 1M rating data sets.}
	\label{fig:accuracy_per_landmarks_1M}
\end{figure*}

As one may observe, MAE decreases as the number of landmarks increases, and the proposed algorithm has outperformed their corresponding \textit{baseline} algorithms with very few landmarks. This behavior was expected, once the more landmarks, the more information are supposedly retained throughout user/item representation.

By comparing landmark selection strategies, \textit{Popularity} has produced the highest accuracies in most of the times. There were few cases in which \textit{Popularity} did not outperform the others as one may observe for user-based CF on MovieLens100k and item-based CF on MovieLens1M. However, in these cases, no other strategy has also consistently outperformed the others.

\textit{Random} and \textit{Coresets Random} have shown the worst performance. Their corresponding MAE values were greater than the other strategies, despite these still remains below the \textit{baseline} algorithms.

We consider that all strategies have performed similarly, especially on MovieLens database. Interestingly, in this database, the MAE difference between landmark selection strategies was very tight, less than $10^{-2}$ for both user-based and item-based CF. From a recommender system perspective, this difference may indicate no greater improvements for the final recommendation list.

One may also note two distinct groups of landmark selection strategies in both \autoref{fig:landmarks_netflix_100k_item-based_mae} and \ref{fig:landmarks_netflix_1M_item-based_mae}. One group uniformly selects landmark at random: \textit{Random} and \textit{Coreset Random}; and another group composed of those strategies that take into account the number of ratings: \textit{Dist. of Ratings}, \textit{Coresets}, and \textit{Popularity}. The difference in accuracy between these groups was higher for few landmarks. This leads us to claim that it is preferred to choose landmarks with more ratings, thus one obtains more co-rated items, and consequently, more `representativeness' in the new user space.

In Tables\footnote[1]{The best result for each landmark selection strategy is highlighted in bold and is marked with an asterisk (`*'). The best result overall is also in bold but marked with double asterisk (`**').} \ref{tab:landmarks_ml_100k_distance_mae}, \ref{tab:landmarks_netflix_100k_distance_mae}, \ref{tab:landmarks_ml_1M_distance_mae} and \ref{tab:landmarks_netflix_1M_distance_mae}, we present the results obtained by the proposed method with different combinations of similarity measures to build both the user-landmark matrix and the user-user similarity matrix. There were evaluated three distance measures: Euclidean, Cosine and Pearson. We fixed the number of landmarks in 20 for MovieLens data sets, and 30 for Netflix.

\sisetup{round-mode=places,detect-weight=true}

\robustify\bfseries

\begin{table*}[t]
	\fontsize{7pt}{10pt}
	\selectfont
    \caption{MAE of the user/item-based CF on MovieLens100k. `UCF' and `ICF' stand for User-based CF and Item-based CF, respectively.}
    \centering
    \begin{tabular}{l l S[table-format=1.3,round-precision=3] S[table-format=1.3,round-precision=3] S[table-format=1.3,round-precision=3] S[table-format=1.3,round-precision=3] S[table-format=1.3,round-precision=3] S[table-format=1.3,round-precision=3] S[table-format=1.3,round-precision=3] S[table-format=1.3,round-precision=3] S[table-format=1.3,round-precision=3] S[table-format=1.3,round-precision=3]}
    \hline\noalign{\smallskip}
    \multirow{2}{*}{Users-Landmarks} & \multirow{2}{*}{User-User} & \multicolumn{2}{l}{Random} & \multicolumn{2}{l}{Dist. of Ratings} & \multicolumn{2}{l}{Coresets} & \multicolumn{2}{l}{Coresets Random} & \multicolumn{2}{l}{Popularity} \\
    \cline{3-12}
    & & {UCF} & {ICF} & {UCF} & {ICF} & {UCF} & {ICF} & {UCF} & {ICF} & {UCF} & {ICF} \\
	\noalign{\smallskip}\hline\noalign{\smallskip}
    \multirow{3}{*}{Euclidean}	&	Euclidean	&	0.72496	&	\bfseries 0.70802{$^{*}$}	&	0.72412	&	0.70535	&	0.72399	&	0.70506	&	0.72277	&	\bfseries 0.71058{$^{*}$}	&	0.72285	&	0.70610 \\
    &	Cosine	&	\bfseries 0.71875{$^{*}$}	&	0.70978	&	0.71792	&	0.70396	&	0.71730	&	0.70356	&	0.71918	&	0.71481	&	0.71455	&	0.70205 \\
    &	Pearson	&	0.72092	&	0.71114	&	0.71660	&	0.70535	&	0.71368	&	0.70419	&	0.71813	&	0.71498	&	0.71287	&	0.70002 \\
    \hline
    \multirow{3}{*}{Cosine}	&	Euclidean	&	0.72435	&	0.71543	&	0.71476	&	\bfseries 0.70311{$^{*}$}	&	0.71008	&	\bfseries 0.69757{$^{*}$}	&	\bfseries 0.71607{$^{*}$}	&	0.71703	&	0.70502	&	0.68967 \\
    &	Cosine	&	0.72036	&	0.71746	&	\bfseries 0.71108{$^{*}$}	&	0.70472	&	\bfseries 0.70738{$^{*}$}	&	0.69890	&	0.71902	&	0.71700	&	\bfseries 0.70141{$^{**}$}	&	0.69017 \\
    &	Pearson	&	\bfseries 0.71852{$^{*}$}	&	0.71488	&	0.71275	&	0.70478	&	0.70785	&	0.70178	&	0.71913	&	0.71817	&	0.70275	&	\bfseries 0.68824{$^{**}$} \\
    \hline
    \multirow{3}{*}{Pearson}	&	Euclidean	&	0.72310	&	0.72064	&	0.71592	&	0.71051	&	0.70923	&	0.70424	&	0.72067	&	0.71829	&	0.70280	&	0.69582 \\
    &	Cosine	&	0.72212	&	0.71804	&	0.71466	&	0.70724	&	0.71222	&	0.70304	&	0.72192	&	0.72504	&	0.70345	&	0.69570 \\
    &	Pearson	&	0.72802	&	0.71665	&	0.72198	&	0.71107	&	0.71616	&	0.70766	&	0.72919	&	0.72604	&	0.70887	&	0.69792 \\
    \noalign{\smallskip}\hline
    \end{tabular}
    \label{tab:landmarks_ml_100k_distance_mae}
\end{table*}
\sisetup{round-mode=places,detect-weight=true}

\robustify\bfseries

\begin{table*}[t]
	\fontsize{7pt}{10pt}
	\selectfont
    \caption{MAE of the user/item-based CF on Netflix100k. `UCF' and `ICF' stand for User-based CF and Item-based CF, respectively.}
    \centering
    \begin{tabular}{l l S[table-format=1.3,round-precision=3] S[table-format=1.3,round-precision=3] S[table-format=1.3,round-precision=3] S[table-format=1.3,round-precision=3] S[table-format=1.3,round-precision=3] S[table-format=1.3,round-precision=3] S[table-format=1.3,round-precision=3] S[table-format=1.3,round-precision=3] S[table-format=1.3,round-precision=3] S[table-format=1.3,round-precision=3]}
    \hline\noalign{\smallskip}
    \multirow{2}{*}{Users-Landmarks} & \multirow{2}{*}{User-User} & \multicolumn{2}{l}{Random} & \multicolumn{2}{l}{Dist. of Ratings} & \multicolumn{2}{l}{Coresets} & \multicolumn{2}{l}{Coresets Random} & \multicolumn{2}{l}{Popularity} \\
    \cline{3-12}
    & & {UCF} & {ICF} & {UCF} & {ICF} & {UCF} & {ICF} & {UCF} & {ICF} & {UCF} & {ICF} \\
    \noalign{\smallskip}\hline\noalign{\smallskip}
    \multirow{3}{*}{Euclidean}	&	Euclidean	&	0.78572	&	\bfseries 0.77027$^{*}$	&	0.78390	&	0.76676	&	0.78501	&	0.76590	&	0.78685	&	\bfseries 0.77070$^{*}$	&	0.78615	&	0.76440	\\
    &	Cosine	&	0.78403	&	0.77068	&	0.77443	&	0.75841	&	0.77554	&	0.75775	&	0.77948	&	0.77837	&	0.77012	&	0.75366	\\
    &	Pearson	&	0.78419	&	0.77757	&	0.77573	&	0.76069	&	0.77594	&	0.75985	&	0.78143	&	0.77664	&	0.77001	&	0.74644	\\
    \hline
    \multirow{3}{*}{Cosine}	&	Euclidean	&	0.77647	&	0.77335	&	\bfseries 0.76294{$^{*}$}	&	\bfseries 0.75043$^{*}$	&	0.76133	&	\bfseries 0.74349$^{*}$	&	0.77395	&	0.77858	&	0.75680	&	0.73969	\\
    &	Cosine	&	0.77480	&	0.77440	&	0.76434	&	0.75446	&	0.75929	&	0.74738	&	\bfseries 0.77197{$^{*}$}	&	0.77886	&	\bfseries 0.75181{$^{**}$}	&	0.73907	\\
    &	Pearson	&	\bfseries 0.77370{$^{*}$}	&	0.77482	&	0.76706	&	0.75315	&	\bfseries 0.75812{$^{*}$}	&	0.74192	&	0.77569	&	0.78212	&	0.75478	&	\bfseries 0.72998$^{**}$	\\
    \hline
    \multirow{3}{*}{Pearson}	&	Euclidean	&	0.78134	&	0.78404	&	0.77739	&	0.76080	&	0.76855	&	0.75210	&	0.77876	&	0.78580	&	0.76261	&	0.74505	\\
    &	Cosine	&	0.78659	&	0.78433	&	0.77748	&	0.76234	&	0.77054	&	0.75052	&	0.78345	&	0.78810	&	0.76750	&	0.74270	\\
    &	Pearson	&	0.79138	&	0.78634	&	0.78229	&	0.76252	&	0.78036	&	0.75334	&	0.78667	&	0.79357	&	0.77499	&	0.74670	\\
	\noalign{\smallskip}\hline
    \end{tabular}
    \label{tab:landmarks_netflix_100k_distance_mae}
\end{table*}
\sisetup{round-mode=places,detect-weight=true}

\robustify\bfseries

\begin{table*}[t]
    \fontsize{7pt}{10pt}
    \selectfont
    \caption{MAE of the user/item-based CF on MovieLens1M. `UCF' and `ICF' stand for User-based CF and Item-based CF, respectively.}
    \centering
    \begin{tabular}{l l S[table-format=1.3,round-precision=3] S[table-format=1.3,round-precision=3] S[table-format=1.3,round-precision=3] S[table-format=1.3,round-precision=3] S[table-format=1.3,round-precision=3] S[table-format=1.3,round-precision=3] S[table-format=1.3,round-precision=3] S[table-format=1.3,round-precision=3] S[table-format=1.3,round-precision=3] S[table-format=1.3,round-precision=3]}
   	\hline\noalign{\smallskip}
   	\multirow{2}{*}{Users-Landmarks} & \multirow{2}{*}{User-User} & \multicolumn{2}{l}{Random} & \multicolumn{2}{l}{Dist. of Ratings} & \multicolumn{2}{l}{Coresets} & \multicolumn{2}{l}{Coresets Random} & \multicolumn{2}{l}{Popularity} \\
   	\cline{3-12}
    & & {UCF} & {ICF} & {UCF} & {ICF} & {UCF} & {ICF} & {UCF} & {ICF} & {UCF} & {ICF} \\
    \noalign{\smallskip}\hline\noalign{\smallskip}
    \multirow{3}{*}{Euclidean}	&	Euclidean	&	0.70446	&	0.67022	&	0.70033	&	0.67082	&	0.69911	&	0.67184	&	0.70149	&	0.67127	&	0.69703	&	0.67705	\\
    &	Cosine	&	0.69750	&	\bfseries 0.66558{$^{*}$}	&	0.69147	&	\bfseries 0.66042{$^{*}$}	&	0.68825	&	0.66095	&	0.69404	&	\bfseries 0.66478{$^{*}$}	&	0.68205	&	0.66483	\\
    &	Pearson	&	\bfseries 0.69467{$^{*}$}	&	0.66850	&	0.69088	&	0.66093	&	0.68820	&	0.66223	&	0.69464	&	0.66605	&	0.68055	&	0.66620	\\
    \hline
    \multirow{3}{*}{Cosine}	&	Euclidean	&	0.69845	&	0.67719	&	\bfseries 0.68904{$^{*}$}	&	0.66486	&	0.68353	&	0.66027	&	\bfseries 0.69225{$^{*}$}	&	0.67082	&	0.67562	&	0.65892	\\
    &	Cosine	&	0.69747	&	0.68102	&	\bfseries 0.68949{$^{*}$}	&	0.66450	&	\bfseries 0.68162{$^{*}$}	&	0.65868	&	0.69388	&	0.67484	&	\bfseries 0.67275{$^{**}$}	&	0.65082	\\
    &	Pearson	&	0.70291	&	0.67894	&	0.69037	&	0.66348	&	0.68421	&	\bfseries 0.65569{$^{*}$}	&	0.69525	&	0.67488	&	\bfseries 0.67343{$^{**}$}	&	\bfseries 0.64806{$^{**}$}	\\
    \hline
    \multirow{3}{*}{Pearson}	&	Euclidean	&	0.70625	&	0.68073	&	0.69272	&	0.66987	&	0.68799	&	0.66345	&	0.69819	&	0.67904	&	0.67946	&	0.65840	\\
    &	Cosine	&	0.70387	&	0.68355	&	0.69397	&	0.66836	&	0.68907	&	0.66094	&	0.69780	&	0.67908	&	0.67867	&	0.65534	\\
    &	Pearson	&	0.71200	&	0.68356	&	0.69924	&	0.66780	&	0.69167	&	0.66196	&	0.70514	&	0.68200	&	0.68018	&	0.65707	\\
    \noalign{\smallskip}\hline
    \end{tabular}
    \label{tab:landmarks_ml_1M_distance_mae}
\end{table*}
\sisetup{round-mode=places,detect-weight=true}

\robustify\bfseries

\begin{table*}[t]
	\fontsize{7pt}{10pt}
	\selectfont
    \caption{MAE of the user/item-based CF on Netflix1M. `UCF' and `ICF' stand for User-based CF and Item-based CF, respectively.}
    \centering
    \begin{tabular}{l l S[table-format=1.3,round-precision=3] S[table-format=1.3,round-precision=3] S[table-format=1.3,round-precision=3] S[table-format=1.3,round-precision=3] S[table-format=1.3,round-precision=3] S[table-format=1.3,round-precision=3] S[table-format=1.3,round-precision=3] S[table-format=1.3,round-precision=3] S[table-format=1.3,round-precision=3] S[table-format=1.3,round-precision=3]}
   	\hline\noalign{\smallskip}
   	\multirow{2}{*}{Users-Landmarks} & \multirow{2}{*}{User-User} & \multicolumn{2}{l}{Random} & \multicolumn{2}{l}{Dist. of Ratings} & \multicolumn{2}{l}{Coresets} & \multicolumn{2}{l}{Coresets Random} & \multicolumn{2}{l}{Popularity} \\
   	\cline{3-12}
    & & {UCF} & {ICF} & {UCF} & {ICF} & {UCF} & {ICF} & {UCF} & {ICF} & {UCF} & {ICF} \\
   	\noalign{\smallskip}\hline\noalign{\smallskip}
	\multirow{3}{*}{Euclidean}	&	Euclidean	&	0.74086	&	0.73437	&	0.73823	&	0.73032	&	0.73650	&	0.72955	&	0.73986	&	0.73548	&	0.73149	&	0.72847	\\
	&	Cosine	&	0.73819	&	\bfseries 0.72742{$^{*}$}	&	0.72965	&	0.71604	&	0.72813	&	0.71382	&	0.73793	&	0.73026	&	0.71873	&	0.71099	\\
	&	Pearson	&	0.73922	&	0.72756	&	0.73113	&	0.71534	&	0.72621	&	0.71292	&	0.73866	&	\bfseries 0.72792{$^{*}$}	&	0.71748	&	0.70874	\\
	\hline
	\multirow{3}{*}{Cosine}	&	Euclidean	&	0.73602	&	0.73772	&	0.72343	&	0.70539	&	0.71716	&	0.70230	&	\bfseries 0.72894{$^{*}$}	&	0.73951	&	0.71126	&	0.70059	\\
	&	Cosine	&	0.73596	&	0.73666	&	\bfseries 0.72003{$^{*}$}	&	\bfseries 0.70077{$^{*}$}	&	\bfseries 0.71497{$^{*}$}	&	0.69570	&	\bfseries 0.72886{$^{*}$}	&	0.73749	&	\bfseries 0.70834{$^{**}$}	&	0.69323	\\
	&	Pearson	&	\bfseries 0.73348{$^{*}$}	&	0.73143	&	0.72179	&	\bfseries 0.70101{$^{*}$}	&	0.71602	&	\bfseries 0.69474{$^{*}$}	&	0.73119	&	0.73495	&	0.70953	&	\bfseries 0.69148{$^{**}$}	\\
	\hline
	\multirow{3}{*}{Pearson}	&	Euclidean	&	0.73879	&	0.73795	&	0.72758	&	0.70428	&	0.72099	&	0.69883	&	0.73438	&	0.73989	&	0.71365	&	0.69662	\\
	&	Cosine	&	0.74162	&	0.73434	&	0.72754	&	0.70519	&	0.72238	&	0.69581	&	0.73669	&	0.74027	&	0.71368	&	0.69208	\\
	&	Pearson	&	0.74451	&	0.73986	&	0.73294	&	0.70724	&	0.72518	&	0.69953	&	0.74092	&	0.73834	&	0.71580	&	0.69654	\\
	\noalign{\smallskip}\hline
    \end{tabular}
    \label{tab:landmarks_netflix_1M_distance_mae}
\end{table*}

One should note that, \textit{Popularity} with Cosine distance to build both matrices (user-landmark and user-user) in user-based CF has achieved the best accuracy overall. In item-based CF, the best measure combination was Cosine and Pearson distances to build item-landmark and item-item matrices, respectively. Nevertheless, \textit{Popularity} has performed with relatively similar accuracy to the other combinations of similarity measures. This similar behavior holds for other selection strategies by differing in MAE about $10^{-2}$ and $10^{-3}$, what is insignificant.

Accordingly, we conclude that accuracy increases with the number of landmarks, the most accurate landmark selection strategies are those based on rating distribution, and the choice of similarity measures does not bring significant advantages for accuracy. Additionally, it is possible to outperform the corresponding \textit{baseline} algorithms with very few landmarks -- 10 to 40 landmarks quickly improve accuracy.

\subsubsection{Computational Performance Analysis}

\autoref{tab:landmarks_ml_100k_runtime}, \ref{tab:landmarks_netflix_100k_runtime}, \ref{tab:landmarks_ml_1M_runtime} and \ref{tab:landmarks_netflix_1M_runtime} show the corresponding time required by the proposed algorithm for different parameter settings and data sets. The idea is to investigate the impact on the computational performance with regards to the number of landmarks, the distance measures, and the selection strategies. Thus, we consider the runtime in seconds to build the similarity matrix and to compute rating predictions for the test set.

\sisetup{round-mode=places}

\begin{table*}[t]
    \fontsize{8pt}{10pt}
    \selectfont
    \caption{User/Item-based CF runtime, in seconds, on MovieLens100k. `UCF' and `ICF' stand for User-based CF and Item-based CF, respectively.}
    \centering
    \begin{tabular}{l S[table-format=1.1,round-precision=1] S[table-format=1.1,round-precision=1] S[table-format=1.1,round-precision=1] S[table-format=1.1,round-precision=1] S[table-format=1.1,round-precision=1] S[table-format=1.1,round-precision=1] S[table-format=1.1,round-precision=1] S[table-format=1.1,round-precision=1] S[table-format=1.1,round-precision=1] S[table-format=1.1,round-precision=1]}
    \hline\noalign{\smallskip}
    \multirow{2}{*}{Landmarks} & \multicolumn{2}{l}{Random} & \multicolumn{2}{l}{Dist. of Ratings} & \multicolumn{2}{l}{Coresets} & \multicolumn{2}{l}{Coresets Random} & \multicolumn{2}{l}{Popularity} \\
    \cline{2-11}
    & {UCF} & {ICF} & {UCF} & {ICF} & {UCF} & {ICF} & {UCF} & {ICF} & {UCF} & {ICF} \\
    \noalign{\smallskip}\hline\noalign{\smallskip}
    10	&	0.42409	&	0.7404	&	0.43475	&	0.90033	&	1.46318	&	2.21889	&	1.41031	&	1.89092	&	0.45354	&	0.96837	\\
    20	&	0.79714	&	1.32592	&	0.82043	&	1.61818	&	2.62602	&	3.88083	&	2.56887	&	3.19666	&	0.85452	&	1.79462	\\
    30	&	1.18732	&	2.01846	&	1.21653	&	2.48181	&	3.70072	&	5.67075	&	3.60075	&	4.46961	&	1.25582	&	2.63866	\\
    40	&	1.59205	&	2.70654	&	1.64826	&	3.23553	&	4.74359	&	7.13727	&	4.60901	&	5.80822	&	1.66189	&	3.5053	\\
    50	&	1.96417	&	3.42701	&	2.02304	&	4.0477	&	5.72446	&	8.76695	&	5.5997	&	7.24382	&	2.10084	&	4.36858	\\
    60	&	2.41241	&	4.09591	&	2.48671	&	4.90012	&	6.71988	&	10.28617	&	6.52256	&	8.47232	&	2.55454	&	5.27785	\\
    70	&	2.7942	&	4.84919	&	2.89899	&	5.71783	&	7.78091	&	11.7866	&	7.65472	&	9.535	&	2.98402	&	6.15241	\\
    80	&	3.22035	&	5.515	&	3.34756	&	6.51029	&	8.65535	&	13.27842	&	8.41822	&	10.65086	&	3.41868	&	6.95265	\\
    90	&	3.60958	&	6.19765	&	3.65799	&	7.30541	&	9.54367	&	14.69196	&	9.35196	&	11.74731	&	3.79926	&	7.77993	\\
    100	&	3.97974	&	6.94693	&	4.08487	&	8.06271	&	10.62758	&	16.22485	&	10.15839	&	13.12436	&	4.16694	&	8.60951 \\
    \noalign{\smallskip}\hline
    \end{tabular}
    \label{tab:landmarks_ml_100k_runtime}
\end{table*}
\sisetup{round-mode=places}

\begin{table*}[t]
    \fontsize{8pt}{10pt}
    \selectfont
    \caption{User/Item-based CF runtime, in seconds, on Netflix100k. `UCF' and `ICF' stand for User-based CF and Item-based CF, respectively.}
    \centering
    \begin{tabular}{l S[table-format=1.1,round-precision=1] S[table-format=1.1,round-precision=1] S[table-format=1.1,round-precision=1] S[table-format=1.1,round-precision=1] S[table-format=1.1,round-precision=1] S[table-format=1.1,round-precision=1] S[table-format=1.1,round-precision=1] S[table-format=1.1,round-precision=1] S[table-format=1.1,round-precision=1] S[table-format=1.1,round-precision=1]}
    \hline\noalign{\smallskip}
    \multirow{2}{*}{Landmarks} & \multicolumn{2}{l}{Random} & \multicolumn{2}{l}{Dist. of Ratings} & \multicolumn{2}{l}{Coresets} & \multicolumn{2}{l}{Coresets Random} & \multicolumn{2}{l}{Popularity} \\
    \cline{2-11}
    & {UCF} & {ICF} & {UCF} & {ICF} & {UCF} & {ICF} & {UCF} & {ICF} & {UCF} & {ICF} \\
    \noalign{\smallskip}\hline\noalign{\smallskip}
    10	&	0.86937	&	1.19235	&	0.93364	&	1.53193	&	3.00038	&	3.59334	&	2.82696	&	3.0021	&	0.99159	&	1.74192	\\
    20	&	1.59515	&	2.17892	&	1.76389	&	2.72459	&	5.49266	&	6.52396	&	5.10923	&	5.34648	&	1.84472	&	3.15164	\\
    30	&	2.37513	&	3.23249	&	2.61193	&	4.07302	&	7.98733	&	9.38865	&	7.53925	&	7.6756	&	2.78829	&	4.57098	\\
    40	&	3.17234	&	4.29321	&	3.48798	&	5.44352	&	10.39311	&	12.42767	&	9.72532	&	10.16246	&	3.62345	&	6.12516	\\
    50	&	4.00679	&	5.24847	&	4.36172	&	6.72857	&	12.74122	&	15.15714	&	12.11009	&	12.53102	&	4.53777	&	7.57815	\\
    60	&	4.84173	&	6.5152	&	5.17251	&	8.16645	&	15.16441	&	18.22403	&	14.04662	&	14.77138	&	5.37321	&	9.04216	\\
    70	&	5.51581	&	7.3993	&	6.00406	&	9.40466	&	17.20758	&	20.75704	&	16.27922	&	17.1477	&	6.25473	&	10.54644	\\
    80	&	6.37325	&	8.54399	&	6.81667	&	11.03002	&	19.48492	&	24.55323	&	18.3647	&	19.55651	&	7.2026	&	12.16461	\\
    90	&	7.11238	&	9.66521	&	7.79521	&	12.15478	&	21.67507	&	26.69298	&	20.24884	&	22.0453	&	8.06559	&	13.26284	\\
    100	&	8.03919	&	10.61701	&	8.62099	&	13.26905	&	23.67384	&	29.27995	&	22.39404	&	24.48799	&	8.98335	&	14.78695 \\
    \noalign{\smallskip}\hline
    \end{tabular}
    \label{tab:landmarks_netflix_100k_runtime}
\end{table*}
\sisetup{round-mode=places}

\begin{table*}[t]
    \fontsize{8pt}{10pt}
    \selectfont
    \caption{User/Item-based CF runtime, in seconds, on MovieLens1M. `UCF' and `ICF' stand for User-based CF and Item-based CF, respectively.}
    \centering
    \begin{tabular}{l S[table-format=2.1,round-precision=1] S[table-format=2.1,round-precision=1] S[table-format=2.1,round-precision=1] S[table-format=2.1,round-precision=1] S[table-format=2.1,round-precision=1] S[table-format=2.1,round-precision=1] S[table-format=2.1,round-precision=1] S[table-format=2.1,round-precision=1] S[table-format=2.1,round-precision=1] S[table-format=2.1,round-precision=1]}
    \hline\noalign{\smallskip}
    \multirow{2}{*}{Landmarks} & \multicolumn{2}{l}{Random} & \multicolumn{2}{l}{Dist. of Ratings} & \multicolumn{2}{l}{Coresets} & \multicolumn{2}{l}{Coresets Random} & \multicolumn{2}{l}{Popularity} \\
    \cline{2-11}
    & {UCF} & {ICF} & {UCF} & {ICF} & {UCF} & {ICF} & {UCF} & {ICF} & {UCF} & {ICF} \\
    \noalign{\smallskip}\hline\noalign{\smallskip}
    10	&	14.41	&	7.68	&	15.31	&	8.59	&	31.97	&	23.54	&	30.81	&	21.58	&	15.87	&	8.93	\\
    20	&	27.18	&	14.57	&	28.79	&	16.27	&	60.95	&	46.51	&	59.46	&	42.59	&	29.68	&	16.83	\\
    30	&	41.18	&	21.84	&	42.82	&	24.66	&	92.43	&	69.42	&	88.22	&	64.86	&	44.66	&	25.49	\\
    40	&	54.82	&	30.00	&	57.19	&	32.89	&	122.97	&	92.68	&	119.93	&	84.83	&	59.68	&	33.86	\\
    50	&	68.47	&	36.88	&	71.96	&	41.02	&	152.70	&	114.39	&	148.50	&	105.59	&	75.62	&	42.22	\\
    60	&	82.11	&	45.22	&	88.16	&	49.15	&	185.16	&	136.22	&	176.00	&	125.04	&	88.58	&	51.81	\\
    70	&	95.30	&	53.34	&	99.86	&	58.00	&	210.74	&	157.04	&	203.18	&	143.53	&	102.55	&	59.64	\\
    80	&	107.91	&	60.73	&	114.33	&	66.49	&	241.25	&	178.62	&	233.05	&	163.04	&	117.74	&	67.94	\\
    90	&	126.24	&	68.32	&	131.34	&	74.47	&	271.69	&	199.02	&	261.23	&	179.68	&	133.99	&	76.40	\\
    100	&	136.78	&	75.45	&	142.75	&	82.57	&	295.33	&	220.14	&	286.09	&	200.74	&	147.26	&	84.50	\\
    \noalign{\smallskip}\hline
    \end{tabular}
    \label{tab:landmarks_ml_1M_runtime}
\end{table*}
\sisetup{round-mode=places}

\begin{table*}[t]
    \fontsize{8pt}{10pt}
    \selectfont
    \caption{User/Item-based CF runtime, in seconds, on Netflix1M. `UCF' and `ICF' stand for User-based CF and Item-based CF, respectively.}
    \centering
    \begin{tabular}{l S[table-format=2.1,round-precision=1] S[table-format=2.1,round-precision=1] S[table-format=2.1,round-precision=1] S[table-format=2.1,round-precision=1] S[table-format=2.1,round-precision=1] S[table-format=2.1,round-precision=1] S[table-format=2.1,round-precision=1] S[table-format=2.1,round-precision=1] S[table-format=2.1,round-precision=1] S[table-format=2.1,round-precision=1]}
    \hline\noalign{\smallskip}
    \multirow{2}{*}{Landmarks} & \multicolumn{2}{l}{Random} & \multicolumn{2}{l}{Dist. of Ratings} & \multicolumn{2}{l}{Coresets} & \multicolumn{2}{l}{Coresets Random} & \multicolumn{2}{l}{Popularity} \\
    \cline{2-11}
    & {UCF} & {ICF} & {UCF} & {ICF} & {UCF} & {ICF} & {UCF} & {ICF} & {UCF} & {ICF} \\
    \noalign{\smallskip}\hline\noalign{\smallskip}
    10	&	26.81	&	9.82	&	29.75	&	11.21	&	58.87	&	35.63	&	56.03	&	29.99	&	31.07	&	11.96	\\
    20	&	50.39	&	18.05	&	55.49	&	21.56	&	113.22	&	69.17	&	105.47	&	60.49	&	57.72	&	23.44	\\
    30	&	74.92	&	27.17	&	82.15	&	32.20	&	170.90	&	103.32	&	160.06	&	88.40	&	86.67	&	33.78	\\
    40	&	103.03	&	36.17	&	110.66	&	42.57	&	227.24	&	137.34	&	216.78	&	118.23	&	114.07	&	44.96	\\
    50	&	127.49	&	45.39	&	138.64	&	53.49	&	285.61	&	165.39	&	268.69	&	144.24	&	143.71	&	54.86	\\
    60	&	153.06	&	54.33	&	169.29	&	63.87	&	340.23	&	197.88	&	316.27	&	173.13	&	170.57	&	66.35	\\
    70	&	177.69	&	63.93	&	194.74	&	74.17	&	393.97	&	231.80	&	366.63	&	204.93	&	199.77	&	79.76	\\
    80	&	201.46	&	74.17	&	220.86	&	86.23	&	447.91	&	263.51	&	415.64	&	232.53	&	231.60	&	91.57	\\
    90	&	227.26	&	83.60	&	250.30	&	97.51	&	499.74	&	286.94	&	462.25	&	250.50	&	256.99	&	98.11	\\
    100	&	253.85	&	89.81	&	277.04	&	104.29	&	552.73	&	313.52	&	508.61	&	278.57	&	286.05	&	111.55	\\
        \hline
    \end{tabular}
    \label{tab:landmarks_netflix_1M_runtime}
\end{table*}

As one would expect, the time increases almost linearly with the number of landmarks. As the dimensionality of user-landmark matrix grows, more computations are necessary to calculate the user-user similarity matrix. Analogously, the same behavior is observed for item-based CF.

In terms of landmark selection strategy, one should note that the simpler strategies outperform the more complex ones. \textit{Random} was away the fastest selection strategy, followed in order by \textit{Dist. of Ratings}, \textit{Popularity}, \textit{Coresets Random}, and \textit{Coresets}. 

The times spent by the \textit{baseline} algorithms are presented in \autoref{tab:baselines_user-based_item-based_time}. Thus, our proposal may achieve a reduction up to 99.22\% in terms of runtime.

\sisetup{round-mode=places}

\begin{table}
    \fontsize{8pt}{10pt}
    \selectfont
	\caption{The runtime, in seconds, of user/item-based CF with Cosine in databases.}
	\centering
	\begin{tabular}{l l S[table-format=1.1,round-precision=1] S[table-format=1.1,round-precision=1] S[table-format=1.1,round-precision=1] S[table-format=1.1,round-precision=1]}
		\hline\noalign{\smallskip}
		 & {CF Type} & {MovieLens100k} & {Netflix100k} & {MovieLens1M} & {MovieLens1M} \\
		\noalign{\smallskip}\hline\noalign{\smallskip}
		\multirow{2}{*}{Time (s)} & User-based & 7.94004 & 24.32867 & 1250.85025 & 3219.83751 \\
		 & Item-based & 15.23886 & 42.56476 & 758.24146 & 1260.00463 \\
		\noalign{\smallskip}\hline
	\end{tabular}
	\label{tab:baselines_user-based_item-based_time}
\end{table}

\textit{Random}, \textit{Dist. of Ratings} and \textit{Popularity} have performed faster than the corresponding \textit{baseline} algorithms for any number of landmarks between 10 and 100 on both data sets of 100k ratings. \textit{Coresets} and \textit{Coresets Random} have made the proposed algorithm slower due to their own complexities. Consequently, the proposal becomes slower than the \textit{baseline} algorithms after 80 landmarks on MovieLens100k for user-based CF, and after adding 100 landmarks for item-based CF.

Interestingly, all \textit{baseline} algorithms on 1M-sized data set have taken more time to perform than the proposed algorithm for any parameter setting (number of landmarks and selection strategy). Thus, the proposal may succeed well on very large databases.

Tables\footnotemark[1] \ref{tab:landmarks_ml_100k_runtime_distances}, \ref{tab:landmarks_netflix_100k_runtime_distances}, \ref{tab:landmarks_ml_1M_distance_runtime} and \ref{tab:landmarks_netflix_1M_distance_runtime} present the computational performance with regards to different distance measures for similarity.  To build user-landmark and user-user matrices, the fastest distance combinations were Euclidean-Euclidean and Cosine-Euclidean. The same behavior can be observed for the item-based CF. Person has presented the worst performance due to its nonlinear computational complexity, which becomes worse on the data sets of 1M ratings.

\sisetup{round-mode=places,detect-weight=true}

\robustify\bfseries

\begin{table*}[t]
	\fontsize{7pt}{10pt}
	\selectfont
	\caption{User/Item-based runtime, in seconds, on MovieLens100k. `UCF' and `ICF' stand for User-based CF and Item-based CF, respectively.}
	\centering
    	\begin{tabular}{l l S[table-format=1.2,round-precision=2] S[table-format=1.2,round-precision=2] S[table-format=1.2,round-precision=2] S[table-format=1.2,round-precision=2] S[table-format=1.2,round-precision=2] S[table-format=1.2,round-precision=2] S[table-format=1.2,round-precision=2] S[table-format=1.2,round-precision=2] S[table-format=1.2,round-precision=2] S[table-format=1.2,round-precision=2]}
    	\hline\noalign{\smallskip}
    	\multirow{2}{*}{Users-Landmarks} & \multirow{2}{*}{User-User} & \multicolumn{2}{l}{Random} & \multicolumn{2}{l}{Dist. of Ratings} & \multicolumn{2}{l}{Coresets} & \multicolumn{2}{l}{Coresets Random} & \multicolumn{2}{l}{Popularity} \\
    	\cline{3-12}
        & & {UCF} & {ICF} & {UCF} & {ICF} & {UCF} & {ICF} & {UCF} & {ICF} & {UCF} & {ICF} \\
    	\noalign{\smallskip}\hline\noalign{\smallskip}
    	\multirow{3}{*}{Euclidean}	&	Euclidean	&	0.70304	&	\bfseries 1.19348$^{**}$	&	0.71224	&	\bfseries 1.39309$^{*}$	&	2.58878	&	3.65508	&	2.49819	&	3.07948	&	0.75305	&	\bfseries 1.46573$^{*}$	\\
    	&	Cosine	&	0.82390	&	1.37117	&	0.85575	&	1.63950	&	2.69285	&	3.96102	&	2.55726	&	3.14760	&	0.86035	&	1.78585	\\
    	&	Pearson	&	1.19584	&	2.02543	&	1.25327	&	2.54571	&	3.16979	&	4.87957	&	3.02458	&	3.76887	&	1.30344	&	2.82376	\\
    	\hline
    	\multirow{3}{*}{Cosine}	&	Euclidean	&	\bfseries 0.67126$^{**}$	&	1.20520	&	\bfseries 0.69803$^{*}$	&	\bfseries 1.38829$^{*}$	&	\bfseries 2.28129$^{*}$	&	\bfseries 3.35116$^{*}$	&	\bfseries 2.19695$^{*}$	&	\bfseries 2.81323$^{*}$	&	\bfseries 0.71325$^{*}$	&	\bfseries 1.46598$^{*}$	\\
    	&	Cosine	&	0.78216	&	1.31618	&	0.82218	&	1.64228	&	2.39980	&	3.63140	&	2.30990	&	2.85525	&	0.84393	&	1.80910	\\
    	&	Pearson	&	1.16319	&	2.04931	&	1.21199	&	2.55912	&	2.80830	&	4.56189	&	2.71675	&	3.53923	&	1.24987	&	2.81486	\\
    	\hline
    	\multirow{3}{*}{Pearson}	&	Euclidean	&	1.01340	&	1.38831	&	1.03116	&	1.66687	&	2.88357	&	3.90457	&	2.78105	&	3.23539	&	1.08444	&	1.80861	\\
    	&	Cosine	&	1.11055	&	1.52790	&	1.19469	&	1.89856	&	3.06779	&	4.19273	&	2.94522	&	3.33192	&	1.24408	&	2.10646	\\
    	&	Pearson	&	1.48500	&	2.16222	&	1.59385	&	2.76060	&	3.46607	&	5.13553	&	3.29458	&	3.86350	&	1.67914	&	3.07585	\\
    	\noalign{\smallskip}\hline		
	\end{tabular}
	\label{tab:landmarks_ml_100k_runtime_distances}
\end{table*}
\sisetup{round-mode=places,detect-weight=true}

\robustify\bfseries

\begin{table*}[t]
    \fontsize{7pt}{10pt}
    \selectfont
    \caption{User/Item-based CF runtime, in seconds, on Netflix100k. `UCF' and `ICF' stand for User-based CF and Item-based CF, respectively.}
    \centering
    \begin{tabular}{l l S[table-format=1.2,round-precision=2] S[table-format=1.2,round-precision=2] S[table-format=1.2,round-precision=2] S[table-format=1.2,round-precision=2] S[table-format=1.2,round-precision=2] S[table-format=1.2,round-precision=2] S[table-format=1.2,round-precision=2] S[table-format=1.2,round-precision=2] S[table-format=1.2,round-precision=2] S[table-format=1.2,round-precision=2]}
    	\hline\noalign{\smallskip}
    	\multirow{2}{*}{Users-Landmarks} & \multirow{2}{*}{User-User} & \multicolumn{2}{l}{Random} & \multicolumn{2}{l}{Dist. of Ratings} & \multicolumn{2}{l}{Coresets} & \multicolumn{2}{l}{Coresets Random} & \multicolumn{2}{l}{Popularity} \\
    	\cline{3-12}
        & & {UCF} & {ICF} & {UCF} & {ICF} & {UCF} & {ICF} & {UCF} & {ICF} & {UCF} & {ICF} \\
	    \noalign{\smallskip}\hline\noalign{\smallskip}
		 \multirow{3}{*}{Euclidean}	&	Euclidean	&	\bfseries 1.44079$^{**}$	&	\bfseries  2.92349$^{**}$	&	\bfseries 1.55418$^{*}$	&	\bfseries  3.46116$^{*}$	&	\bfseries 5.38843$^{*}$	&	8.60638	&	\bfseries 5.07785$^{*}$	&	7.33873	&	\bfseries 1.58848$^{*}$	&	\bfseries  3.78230$^{*}$	\\
		 &	Cosine	&	1.64648	&	3.06778	&	1.84462	&	3.87909	&	5.63716	&	9.20368	&	5.30732	&	7.36493	&	1.91706	&	4.44964	\\
		 &	Pearson	&	2.35041	&	4.10959	&	2.67240	&	5.83786	&	6.60672	&	11.14741	&	5.99361	&	8.47749	&	2.80932	&	6.73469	\\
		 \hline
		 \multirow{3}{*}{Cosine}	&	Euclidean	&	2.27070	&	2.95663	&	2.38217	&	3.50500	&	7.65956	&	\bfseries 8.59575$^{*}$	&	7.17357	&	\bfseries 7.23595$^{*}$	&	2.47214	&	3.79993	\\
		 &	Cosine	&	2.52077	&	3.07672	&	2.79426	&	3.98113	&	7.89586	&	9.12172	&	7.31378	&	7.45273	&	2.93868	&	4.49466	\\
		 &	Pearson	&	3.70559	&	4.09585	&	4.10699	&	5.83150	&	9.30182	&	11.24297	&	8.60354	&	8.30926	&	4.28549	&	6.73914	\\
		 \hline
		 \multirow{3}{*}{Pearson}	&	Euclidean	&	1.93902	&	3.30470	&	2.14911	&	4.11223	&	5.96879	&	9.29703	&	5.57287	&	7.66596	&	2.26514	&	4.56460	\\
		 &	Cosine	&	2.11469	&	3.41740	&	2.41919	&	4.38877	&	6.30514	&	9.67810	&	5.93072	&	7.67387	&	2.58669	&	5.02878	\\
		 &	Pearson	&	2.77170	&	4.15283	&	3.18684	&	5.90292	&	7.09468	&	11.32402	&	6.47632	&	8.46133	&	3.48087	&	6.79830	\\
		 \noalign{\smallskip}\hline
    \end{tabular}
    \label{tab:landmarks_netflix_100k_runtime_distances}
\end{table*}
\sisetup{round-mode=places,detect-weight=true}

\robustify\bfseries

\begin{table*}[t]
    \fontsize{7pt}{10pt}
    \selectfont
    \caption{User/Item-based CF runtime, in seconds, on MovieLens1M. `UCF' and `ICF' stand for User-based CF and Item-based CF, respectively.}
    \centering
    \begin{tabular}{l l S[table-format=1.2,round-precision=2] S[table-format=1.2,round-precision=2] S[table-format=1.2,round-precision=2] S[table-format=1.2,round-precision=2] S[table-format=1.2,round-precision=2] S[table-format=1.2,round-precision=2] S[table-format=1.2,round-precision=2] S[table-format=1.2,round-precision=2] S[table-format=1.2,round-precision=2] S[table-format=1.2,round-precision=2]}
   	\hline\noalign{\smallskip}
   	\multirow{2}{*}{Users-Landmarks} & \multirow{2}{*}{User-User} & \multicolumn{2}{l}{Random} & \multicolumn{2}{l}{Dist. of Ratings} & \multicolumn{2}{l}{Coresets} & \multicolumn{2}{l}{Coresets Random} & \multicolumn{2}{l}{Popularity} \\
   	\cline{3-12}
    & & {UCF} & {ICF} & {UCF} & {ICF} & {UCF} & {ICF} & {UCF} & {ICF} & {UCF} & {ICF} \\
   	\noalign{\smallskip}\hline\noalign{\smallskip}    
    \multirow{3}{*}{Euclidean}	&	Euclidean	&	\bfseries 22.44098$^{*}$	&	13.58327	&	\bfseries 23.42668$^{**}$	&	14.55686	&	55.20740	&	44.98347	&	54.02966	&	41.86817	&	\bfseries 23.68079$^{*}$	&	14.80649	\\
    &	Cosine	&	26.95426	&	15.28108	&	28.87726	&	16.69701	&	61.23771	&	47.32432	&	59.00224	&	43.14884	&	29.66043	&	17.08854	\\
    &	Pearson	&	42.54778	&	19.67283	&	44.79632	&	22.49415	&	77.83162	&	53.14305	&	74.95381	&	48.52080	&	46.42267	&	23.25657	\\
    \hline
    \multirow{3}{*}{Cosine}	&	Euclidean	&	22.93367	&	\bfseries 12.52181$^{**}$	&	23.47330	&	\bfseries 13.51665$^{*}$	&	\bfseries 53.58420$^{*}$	&	\bfseries 40.33578$^{*}$	&	\bfseries 53.66504$^{*}$	&	\bfseries 37.59785$^{*}$	&	24.16343	&	\bfseries 13.59349$^{*}$	\\
    &	Cosine	&	27.87061	&	13.16479	&	29.68623	&	15.01877	&	62.05137	&	41.78053	&	59.87315	&	38.35157	&	29.79747	&	15.56689	\\
    &	Pearson	&	42.16038	&	17.97514	&	45.60322	&	20.50463	&	76.33357	&	47.98794	&	73.77120	&	43.94134	&	47.08989	&	21.46359	\\
    \hline
    \multirow{3}{*}{Pearson}	&	Euclidean	&	29.71378	&	19.11809	&	31.54572	&	21.38121	&	64.24649	&	52.21481	&	62.15878	&	47.79112	&	32.71644	&	22.63680	\\
    &	Cosine	&	33.61282	&	20.13638	&	36.72163	&	23.46291	&	69.49937	&	54.24607	&	66.20428	&	48.47167	&	38.35960	&	24.80486	\\
    &	Pearson	&	48.08717	&	25.52320	&	52.28012	&	29.31093	&	85.78330	&	60.48396	&	81.24377	&	53.61813	&	55.13946	&	30.85329	\\
    \noalign{\smallskip}\hline
    \end{tabular}
    \label{tab:landmarks_ml_1M_distance_runtime}
\end{table*}

\sisetup{round-mode=places,detect-weight=true}

\robustify\bfseries

\begin{table*}[t]
     \fontsize{7pt}{10pt}
    \selectfont
    \caption{User/Item-based CF runtime, in seconds, on Netflix1M. `UCF' and `ICF' stand for User-based CF and Item-based CF, respectively.}
    \centering
    \begin{tabular}{l l S[table-format=2.2,round-precision=2] S[table-format=2.2,round-precision=2] S[table-format=2.2,round-precision=2] S[table-format=2.2,round-precision=2] S[table-format=2.2,round-precision=2] S[table-format=2.2,round-precision=2] S[table-format=2.2,round-precision=2] S[table-format=2.2,round-precision=2] S[table-format=2.2,round-precision=2] S[table-format=2.2,round-precision=2]}
   	\hline\noalign{\smallskip}
   	\multirow{2}{*}{Users-Landmarks} & \multirow{2}{*}{User-User} & \multicolumn{2}{l}{Random} & \multicolumn{2}{l}{Dist. of Ratings} & \multicolumn{2}{l}{Coresets} & \multicolumn{2}{l}{Coresets Random} & \multicolumn{2}{l}{Popularity} \\
   	\cline{3-12}
    & & {UCF} & {ICF} & {UCF} & {ICF} & {UCF} & {ICF} & {UCF} & {ICF} & {UCF} & {ICF} \\
   	\noalign{\smallskip}\hline\noalign{\smallskip}
   	\multirow{3}{*}{Euclidean}	&	Euclidean	&	64.05931	&	\bfseries 25.07026$^{**}$	&	67.68631	&	\bfseries 27.85013$^{*}$	&	153.63045	&	\bfseries 96.30937$^{*}$	&	146.64899	&	\bfseries 84.84519$^{*}$	&	68.62209	&	\bfseries 28.92011$^{*}$	\\
   	&	Cosine	&	75.96421	&	26.21311	&	81.72213	&	30.55605	&	170.04522	&	99.91414	&	159.74350	&	86.23385	&	84.91974	&	32.86977	\\
   	&	Pearson	&	114.98882	&	31.49290	&	126.05337	&	40.07972	&	215.21496	&	109.12883	&	196.13679	&	92.30268	&	132.18290	&	42.84569	\\
   	\hline
   	\multirow{3}{*}{Cosine}	&	Euclidean	&	\bfseries 60.79719$^{**}$	&	25.93149	&	\bfseries 64.15426$^{*}$	&	28.96631	&	\bfseries 145.08028$^{*}$	&	97.35745	&	\bfseries 138.85900$^{*}$	&	85.59548	&	\bfseries 66.03413$^{*}$	&	29.99604	\\
   	&	Cosine	&	71.32978	&	26.40346	&	78.22777	&	31.38679	&	159.54941	&	101.10540	&	148.96944	&	87.00769	&	81.32675	&	33.57884	\\
   	&	Pearson	&	108.42818	&	33.81295	&	121.18939	&	41.10585	&	203.25513	&	110.32023	&	189.93030	&	91.67938	&	126.64013	&	43.98876	\\
   	\hline
   	\multirow{3}{*}{Pearson} &	Euclidean	&	80.55232	&	33.13249	&	87.49140	&	42.01688	&	174.97722	&	111.45401	&	165.95000	&	92.19849	&	91.34720	&	44.93794	\\
   	&	Cosine	&	89.80493	&	34.72680	&	101.65081	&	43.91678	&	189.86494	&	114.40628	&	174.54750	&	92.30261	&	107.27278	&	47.98238	\\
   	&	Pearson	&	126.17847	&	38.95145	&	145.39498	&	52.65014	&	238.83105	&	122.86228	&	219.20219	&	97.15113	&	156.31046	&	57.39919	\\
   	\noalign{\smallskip}\hline
    \end{tabular}
    \label{tab:landmarks_netflix_1M_distance_runtime}
\end{table*}

Therefore, any of these distance measures may be applied to small databases without great influence on the computational performance. Nevertheless, Euclidean and Cosine should be preferred in case of very large databases. It is also interesting to highlight that our proposal has considerably reduced the runtime compared to the \textit{baseline} algorithms, even with Pearson distance.

\subsection{Comparative Analysis}

We here aim at comparing the proposed algorithm in terms of accuracy and performance to CF algorithms of the state-of-the-art. In this experiment, our proposal is set with \textit{Popularity} for landmark selection, and Cosine similarity to build both reduced and similarity matrix. Additionally, one used 20 landmarks for MovieLens data sets and 30 for Netflix. Yet, both user-based and item-based CF were set with 13 neighbors for rating prediction. 

The comparison considers 8 recommender algorithms: kNN with Euclidean, kNN with Cosine, kNN with Pearson, Regularized Singular Value Decomposition (RSVD), Improved Regularized Singular Value Decomposition (IRSVD), Probability Matrix Factorization (PMF), Bayesian Probability Matrix Factorization (BPMF), and SVD++. The proposed algorithm is here referred as \textit{Landmarks kNN}.

\subsubsection{Accuracy Analysis}

Figures \ref{fig:baseline_accuracy_100k} and \ref{fig:baseline_accuracy_1M} show accuracies achieved by the proposal and the other CF algorithms. The horizontal line crossing the boxplot graph indicates the MAE median of 10-fold cross validation obtained with \textit{Landmarks kNN}.

\input{baselines_accuracy_100k_data.tex}
\input{baselines_accuracy_1M_data.tex}

As one should note, most of memory-based CF algorithms have higher MAE than model-based ones, mainly on the data sets of 1M ratings. This was expected, once model-based CF has been often superior in the sense of accuracy.

In spite of \textit{Landmarks kNN} has been classified as memory-based, it outperformed some model-based ones like IRSVD, PMF and SVD++, and yielded higher accuracy for the item-based CF on the data sets of 100k ratings, as one may see in \autoref{fig:baseline_accuracy_100k}. However, this behavior does not stand on the data sets of 1M ratings, wherein model-based CF algorithms have outperformed our proposal.

Among memory-based CF algorithms, Euclidean kNN has achieved the highest MAE, while Cosine kNN and Pearson kNN have reached intermediate values. 

In case of model-based CF approach, BPMF has shown the lowest MAE on the data sets of 100k ratings, while RSVD has outperformed all model-based CF algorithms on the data sets of 1M ratings. 

\autoref{fig:baseline_accuracy_1M} clearly distinguishes the two groups of CF approaches, \textit{i.e.} memory-based and model-based. The first one has provided higher MAE than the proposed method, \textit{Landmarks kNN}. On the other hand, the second group has beaten our proposal with relative margin.

Therefore, we may note that our proposal outperformed memory-based CF algorithms on all data sets. Additionally, it has yielded similar accuracy to some model-based techniques on the data sets of 100k ratings.

\subsubsection{Computational Performance Analysis}

\sisetup{round-mode=places,detect-weight=true}

\robustify\bfseries

\begin{table*}[t]
    \fontsize{8pt}{10pt}
    \selectfont
	\caption{The table presents how many times the corresponding algorithm is slower than the proposal, which appears in bold. For instance, Euclidean kNN is 8.7 times slower than \textit{Landmarks kNN} on MovieLens100k.}
	\centering
	\begin{tabular}{l S[table-format=2.1,round-precision=1] S[table-format=2.1,round-precision=1] l S[table-format=2.1,round-precision=1] S[table-format=2.1,round-precision=1] l S[table-format=2.1,round-precision=1] S[table-format=2.1,round-precision=1] l S[table-format=2.1,round-precision=1] S[table-format=2.1,round-precision=1]}
		\hline\noalign{\smallskip}
		\multirow{2}{*}{CF Technique} & \multicolumn{6}{l}{User-based} & \multicolumn{5}{l}{Item-based} \\
		\cline{2-6}\cline{8-12}
		& \multicolumn{2}{l}{MovieLens} & & \multicolumn{2}{l}{Netflix} & & \multicolumn{2}{l}{MovieLens} & & \multicolumn{2}{l}{Netflix} \\
		\cline{2-3}\cline{5-6}\cline{8-9}\cline{11-12}
		& {100k} & {1M} & & {100k} & {1M} & & {100k} & {1M} & & {100k} & {1M} \\
		\noalign{\smallskip}\hline\noalign{\smallskip}
        Euclidean kNN	&	8.7	&	39.5 &	&	8.9	&	39.2 &	&	8.3	&	44.8	& &	9.1	&	37.6	\\
        Cosine kNN	&	8.8	&	39.7 &	&	9.0	&	39.1 &	&	8.4	&	45.7 &	&	9.2	&	37.8	\\
        Pearson kNN	&	17.1	&	76.3 &	&	15.7	&	70.8 &	&	14.2	&	78.5 &	&	13.1	&	56.1	\\
        Landmarks kNN	&	\bfseries 1.0	&	\bfseries 1.0 &	&	\bfseries 1.0	&	\bfseries 1.0 &	&	\bfseries 1.0	&	\bfseries 1.0 &	&	\bfseries 1.0	&	\bfseries 1.0	\\
        RSVD	&	49.2	&	16.6 &	&	15.8	&	6.6 &	&	23.3	&	33.5 &	&	9.8	&	15.4	\\
        IRSVD	&	70.9	&	23.3 &	&	22.8	&	9.0 &	&	33.9	&	46.2 &	&	13.8	&	21.2	\\
        PMF	&	8.3	&	3.1 &	&	2.8	&	1.19 &	&	4.1	&	6.5 &	&	1.7	&	2.7	\\
        BPMF	&	50.3	&	10.1 &	&	24.2	&	5.0 &	&	24.8	&	19.8 &	&	14.5	&	12.1	\\
        SVD++	&	437.1	&	297.8 &	&	177.9	&	134.0 &	&	161.1	&	828.6 &	&	85.6	&	541.2	\\
        \noalign{\smallskip}\hline
	\end{tabular}
	\label{tab:baselines_100k_and_1M_time}
\end{table*}


\autoref{tab:baselines_100k_and_1M_time} presents the comparative performance among the CF algorithms. The values indicates how many times each CF algorithm is slower than our proposal (in bold). 

By comparing our proposal with memory-based CF algorithms it is at least 8 times faster than the fastest memory-based. This difference becomes smaller when compared with model-based CF algorithms, wherein PMF performs 1.2 times slower than the proposed method on Netflix1M. However, the other model-based CF algorithms are away slower. 

Therefore, we conclude that our proposal consistently and considerably outperformed the compared state-of-the-art CF algorithms, in the sense of computation performance. 

\subsubsection{Compromise Between Accuracy and Runtime}

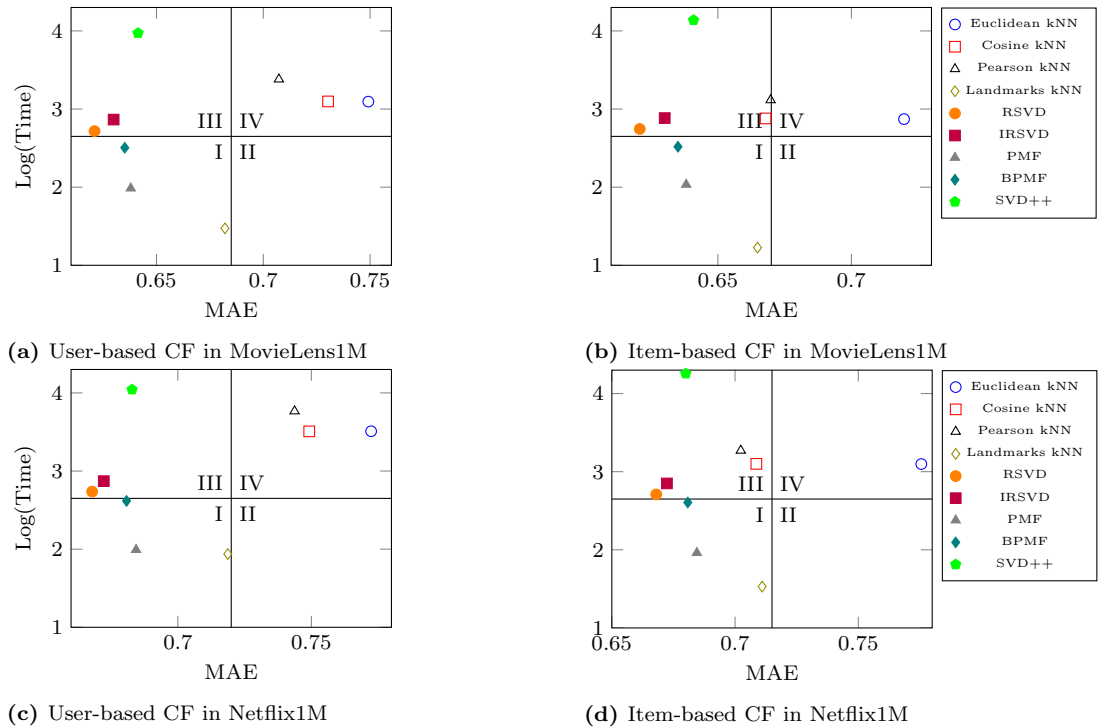
\begin{figure*}
	\begin{subfigure}{0.4\textwidth}
		\begin{tikzpicture}
		\begin{axis}[%
		scatter/classes={%
			Euclidean={mark=o,draw=blue},
			Cosine={mark=square,draw=red},
			Pearson={mark=triangle,draw=black},
			Landmarks={mark=diamond,draw=olive},
			RSVD={mark=*,draw=orange,fill=orange},
			IRSVD={mark=square*,draw=purple,fill=purple},
			PMF={mark=triangle*,draw=gray,fill=gray},
			BPMF={mark=diamond*,draw=teal,fill=teal},
			SVD++={mark=pentagon*,draw=green,fill=green}},
		xlabel=MAE,
		ylabel=Log(Time),
		xmin=0.61,
		xmax=0.76,
		ymin=1.0,
		ymax=4.3,
		legend style={font=\fontsize{4.5}{4.5}\selectfont},
		legend pos=south east,
		height=5cm]
		\addplot[scatter,only marks,%
		scatter src=explicit symbolic]%
		table[meta=label] {
			x	y	label
			0.74919	3.0950156154	Euclidean
			0.73034	3.0972053197	Cosine
			0.70731	3.3807634958	Pearson
			0.68205	1.4724527757	Landmarks
			0.62098	2.7172596175	RSVD
			0.62993	2.8651538001	IRSVD
			0.63795	1.9854154257	PMF
			0.63515	2.5045054144	BPMF
			0.64136	3.9721648207	SVD++
		};
		
		\addplot+[black, no markers] 
		coordinates{(0.685,1) (0.685,4.3)};
		
		\addplot+[black, no markers] 
		coordinates{(0.61,2.65) (0.76,2.65)};
		
		\node at (axis cs:0.685,2.65) [anchor=north east] {I};
		\node at (axis cs:0.685,2.65) [anchor=north west] {II};
		\node at (axis cs:0.685,2.65) [anchor=south east] {III};
		\node at (axis cs:0.685,2.65) [anchor=south west] {IV};
		\end{axis}
		\end{tikzpicture}
		\caption{User-based CF in MovieLens1M}
		\label{fig:scatter_ml_1M_user-based}
	\end{subfigure}
	~~~~~~~~~~~~~~
	\begin{subfigure}{0.4\textwidth}
		\begin{tikzpicture}
		\begin{axis}[%
		scatter/classes={%
			Euclidean={mark=o,draw=blue},
			Cosine={mark=square,draw=red},
			Pearson={mark=triangle,draw=black},
			Landmarks={mark=diamond,draw=olive},
			RSVD={mark=*,draw=orange,fill=orange},
			IRSVD={mark=square*,draw=purple,fill=purple},
			PMF={mark=triangle*,draw=gray,fill=gray},
			BPMF={mark=diamond*,draw=teal,fill=teal},
			SVD++={mark=pentagon*,draw=green,fill=green}},
		xlabel=MAE,
		xmin=0.61,
		xmax=0.73,
		ymin=1.0,
		ymax=4.3,
		legend style={font=\fontsize{4.5}{4.5}\selectfont},
		legend pos=outer north east,
		height=5cm]
		\addplot[scatter,only marks,%
		scatter src=explicit symbolic]%
		table[meta=label] {
			x	y	label
			0.71972	2.8711547901	Euclidean
			0.66788	2.8798075276	Cosine
			0.66974	3.1148747664	Pearson
			0.66483	1.2260727617	Landmarks
			0.62073	2.7450371719	RSVD
			0.63008	2.8844750343	IRSVD
			0.63813	2.0334096005	PMF
			0.63506	2.5175904514	BPMF
			0.64088	4.1384761492	SVD++
		};
		\legend{Euclidean kNN, Cosine kNN, Pearson kNN, Landmarks kNN, RSVD, IRSVD, PMF, BPMF, SVD++};
		
		\addplot+[black, no markers] 
		coordinates{(0.67,1) (0.67,4.3)};
		
		\addplot+[black, no markers] 
		coordinates{(0.61,2.65) (0.73,2.65)};
		
		\node at (axis cs:0.67,2.65) [anchor=north east] {I};
		\node at (axis cs:0.67,2.65) [anchor=north west] {II};
		\node at (axis cs:0.67,2.65) [anchor=south east] {III};
		\node at (axis cs:0.67,2.65) [anchor=south west] {IV};
		\end{axis}
		\end{tikzpicture}
		\caption{Item-based CF in MovieLens1M}
		\label{fig:scatter_ml_1M_item-based}
	\end{subfigure}
    ~
    
	\begin{subfigure}{0.4\textwidth}
		\begin{tikzpicture}
		\begin{axis}[%
		scatter/classes={%
			Euclidean={mark=o,draw=blue},
			Cosine={mark=square,draw=red},
			Pearson={mark=triangle,draw=black},
			Landmarks={mark=diamond,draw=olive},
			RSVD={mark=*,draw=orange,fill=orange},
			IRSVD={mark=square*,draw=purple,fill=purple},
			PMF={mark=triangle*,draw=gray,fill=gray},
			BPMF={mark=diamond*,draw=teal,fill=teal},
			SVD++={mark=pentagon*,draw=green,fill=green}},
		xlabel=MAE,
		ylabel=Log(Time),
		xmin=0.66,
		xmax=0.78,
		ymin=1.0,
		ymax=4.3,
		legend style={font=\fontsize{4.5}{4.5}\selectfont},
		legend pos=south east,
		height=5cm]
		\addplot[scatter,only marks,%
		scatter src=explicit symbolic]%
		table[meta=label] {
			x	y	label
			0.77239	3.5093421029	Euclidean
			0.74921	3.5078339561	Cosine
			0.74372	3.7661675863	Pearson
			0.71873	1.9378473389	Landmarks
			0.66790	2.7368798304	RSVD
			0.67226	2.8706931388	IRSVD
			0.68436	1.9922114498	PMF
			0.68078	2.6169393131	BPMF
			0.68286	4.0430128921	SVD++
		};
		
		\addplot+[black, no markers] 
		coordinates{(0.72,1) (0.72,4.3)};
		
		\addplot+[black, no markers] 
		coordinates{(0.66,2.65) (0.78,2.65)};
		
		\node at (axis cs:0.72,2.65) [anchor=north east] {I};
		\node at (axis cs:0.72,2.65) [anchor=north west] {II};
		\node at (axis cs:0.72,2.65) [anchor=south east] {III};
		\node at (axis cs:0.72,2.65) [anchor=south west] {IV};
		
		\end{axis}
		\end{tikzpicture}
		\caption{User-based CF in Netflix1M}
		\label{fig:scatter_netflix_1M_user-based}
	\end{subfigure}
	~~~~~~~~~~~~~~
	\begin{subfigure}{0.4\textwidth}
		\begin{tikzpicture}
		\begin{axis}[%
		scatter/classes={%
			Euclidean={mark=o,draw=blue},
			Cosine={mark=square,draw=red},
			Pearson={mark=triangle,draw=black},
			Landmarks={mark=diamond,draw=olive},
			RSVD={mark=*,draw=orange,fill=orange},
			IRSVD={mark=square*,draw=purple,fill=purple},
			PMF={mark=triangle*,draw=gray,fill=gray},
			BPMF={mark=diamond*,draw=teal,fill=teal},
			SVD++={mark=pentagon*,draw=green,fill=green}},
		xlabel=MAE,
		xmin=0.65,
		xmax=0.78,
		ymin=1.0,
		ymax=4.3,
		legend style={font=\fontsize{4.5}{4.5}\selectfont},
		legend pos=outer north east,
		height=5cm]
		\addplot[scatter,only marks,%
		scatter src=explicit symbolic]%
		table[meta=label] {
			x	y	label
			0.77558	3.0978104017	Euclidean
			0.70864	3.1003721405	Cosine
			0.70233	3.2712223023	Pearson
			0.71099	1.5286513637	Landmarks
			0.66801	2.7092951449	RSVD
			0.67236	2.8497441797	IRSVD
			0.68450	1.9614143514	PMF
			0.68082	2.6059047967	BPMF
			0.68001	4.2558388009	SVD++
			
		};
		\legend{Euclidean kNN, Cosine kNN, Pearson kNN, Landmarks kNN, RSVD, IRSVD, PMF, BPMF, SVD++};
		
		\addplot+[black, no markers] 
			coordinates{(0.715,1) (0.715,4.3)};
		
		\addplot+[black, no markers] 
			coordinates{(0.65,2.65) (0.78,2.65)};
			
		\node at (axis cs:0.715,2.65) [anchor=north east] {I};
		\node at (axis cs:0.715,2.65) [anchor=north west] {II};
		\node at (axis cs:0.715,2.65) [anchor=south east] {III};
		\node at (axis cs:0.715,2.65) [anchor=south west] {IV};
		
		\end{axis}
		\end{tikzpicture}
		\caption{Item-based CF in Netflix1M}
		\label{fig:scatter_netflix_1M_item-based}
	\end{subfigure}
	\caption{This figure illustrates Accuracy per Time of CF algorithms. Each CF algorithm is represented by a point in the graphs which were divided in four quadrants, in order to better discern algorithm performances. Note that, user/item-based CF with Landmarks (\textit{Landmarks kNN}), PMF and BPMF are located at the first quadrant, which means these perform faster than others and have comparable accuracy. The graphs were generated using 1M ratings data sets.}
	\label{fig:scatter_plots}
\end{figure*}

In order to compare CF algorithms from different standpoints, we plot the accuracies of each algorithm for the data sets of 1M ratings by their corresponding runtime. As one may see in \autoref{fig:scatter_plots}, x and y axes indicates the accuracy (measured with MAE) and the runtime logarithm (in log-seconds), respectively. Each algorithm is therefore represented by a point in 2D space.

Each graph in \autoref{fig:scatter_plots} has been divided into four quadrants so as to better discern the algorithm performances. The first quadrant is the desired one, since it indicates the lowest values for both MAE and runtime, \textit{i.e.} the best compromise between accuracy and computational performance.

Note that, the proposed algorithm, PMF and BPMF are within the first quadrant in all graphs. Additionally, our proposal has performed faster than any other algorithm, including PMF and BPMF.

Regarding to accuracy, one should observe that model-based CF algorithms have clearly yielded higher accuracy than the other techniques based on similarity. Regularized SVD, Improved Regularized SVD and Bayesian Probabilistic MF have performed similarly, in the sense that, their accuracies and computational performances were comparable. The same might be verified for both Cosine and Pearson (Cosine kNN and Pearson kNN, respectively). This behavior is also observed in \autoref{fig:scatter_ml_1M_user-based} and \autoref{fig:scatter_netflix_1M_user-based}, which correspond to user-based CF, and, analogously, in \autoref{fig:scatter_ml_1M_item-based} and \autoref{fig:scatter_netflix_1M_item-based} for item-based CF.

It is remarkable how our proposal consistently and considerably outperforms CF algorithms in computational performance. Furthermore, it provides rating predictions as accurate as any other memory-based techniques, and may offer an interesting compromise between accuracy and computational performance when compared with model-based CF algorithms.

	\section{Conclusion}

In this paper, we presented a proposal to improve memory-based CF computational performance via rating matrix reduction with landmarks. It consists in representing users by their similarities to few preselected users, namely landmarks. Thus, instead of modeling users by rating vectors and building a user-user similarity matrix, we proposed to locate users by their similarities to landmarks, resulting in a new user-user similarity matrix. Thus, small numbers of landmarks leads to great reduction of the rating matrix. Consequently, it decreases the time spent to compute the posterior user-user similarity matrix.

The proposed method has three parameters that influence the new space representation, and consequently the CF algorithm accuracy and runtime. These are 1)the number of landmarks, 2)the distance measure to compute the user-landmark matrix, and 3)the distance measure to compute user-user similarity matrix. After investigating different parameter settings, we found out that accuracy and runtime increase with the number of landmarks. Besides, all evaluated distance measures (Euclidean, Cosine and Pearson) have yielded similar accuracies.

Another important component of the proposed algorithm is how to select landmarks. There were proposed five different ways of selecting landmarks: \textit{Random}, \textit{Dist. of Ratings}, \textit{Coresets}, \textit{Coresets Random}, and \textit{Popularity}. The most accurate strategy was \textit{Popularity}, while \textit{Random} and \textit{Dist. of Ratings} were the fastest ones.

In order to conduct a fair comparison of the proposed algorithm, we selected eight CF algorithms -- both memory-based and model-based approaches -- to compare their performance on MovieLens and Netflix databases. There were considered two different cuts of each database (100k and 1M ratings). We have taken MAE as an accuracy measure, and the runtime in seconds as a computational performance measure.

The results have shown that our proposal has consistently outperformed the other algorithms in terms of time consuming. It has also improved CF scalability, once its runtime increased almost linearly with the number of landmarks. Furthermore, one yielded the highest accuracy overall with regards to memory-based CF algorithms.

Concluding, our proposal offers a very simple and efficient manner to reduce the cost of similarity computations for memory-based CF algorithms by conferring great speedup without loss of accuracy. 

As future work, a theoretical investigation should be addressed so as to determine the number of landmarks that guarantee accuracy bounds. We want to determine lower bounds for the number of landmarks given an approximation error $\epsilon$. Besides, it also important to investigate how landmarks can be used to improve model-based CF algorithms.
	


	\begin{acknowledgements}
		Our thanks to CNPq/CAPES for funding this research.
	\end{acknowledgements}
	
	\bibliographystyle{spbasic}
	\bibliography{bibliography}
	
\end{document}